\documentclass[a4paper,UKenglish,cleveref, autoref, thm-restate]{lipics-v2021}

\pdfoutput=1 
\hideLIPIcs  

\usepackage{soul}
\usepackage[utf8]{inputenc}
\usepackage{algorithm}
\usepackage{algpseudocode}
\usepackage{amsmath}
\usepackage{amssymb}
\usepackage{amsthm}
\usepackage{tikz-cd}
\usepackage[textsize=scriptsize]{todonotes}
\usepackage{xcolor}
\usepackage{tikz}
\usepackage{enumitem}
\usepackage{hyperref}
\usepackage{cleveref}
\usetikzlibrary{cd, calc, shapes, decorations.shapes, decorations.markings, decorations.pathreplacing, shapes.geometric, shapes.symbols,positioning,arrows}
\usepackage{pgfplots}
\usepackage{float}
\usepackage{url}
\usepackage[normalem]{ulem} 
\usepackage{appendix}
\usepackage{caption}
\usepackage{subcaption}
\usepackage{graphicx}
\usepackage{thm-restate}
\usepackage{comment}
\usepackage{enumitem}
\newcommand{\subscript}[2]{$#1 _ #2$}

\newcommand{\closure}[1]{\overline{#1}}
\newcommand{\lub}{\mathrm{l.u.b.}}

\newcommand{\RR}{\mathbb{R}}

\newcommand{\F}{\mathbb{F}}

\tikzset{twosimp/.style={fill opacity=0.6,fill=gray,draw opacity=0.9}}

\DeclareMathOperator{\push}{push}
\DeclareMathOperator{\dgm}{dgm}

\newtheorem*{property*}{Property}

\title{Switch points of bi-persistence matching distance  } 

\author{Robyn Brooks\footnote{Corresponding author}}{University of Utah, Utah, United States \and \url{https://sites.google.com/view/robynkayebrooks }}{robyn.brooks@utah.edu}{https://orcid.org/0000-0002-1825-0097}{}

\author{Celia Hacker}{Max Planck Institute for Mathematics in the Sciences, Leipzig, Germany }{celia.hacker@mis.mpg.de}{https://orcid.org/0000-0002-1825-0097}{}

\author{Claudia Landi}{DISMI, Universit\`a di Modena e Reggio Emilia, Italy  \and \url{http://personale.unimore.it/rubrica/dettaglio/clandi} } {claudia.landi@unimore.it}{https://orcid.org/0000-0001-8725-4844}{}

\author{Barbara I. Mahler}{KTH Royal Institute of Technology Stockholm, Sweden \and \url{https://www.kth.se/profile/bmahler} }{bmahler@kth.se}{https://orcid.org/0000-0002-1825-0097}{}

\author{Elizabeth R. Stephenson}{Institute of Science and Technology Austria, Austria \and \url{https://www.elizabethrstephenson.com}}{elizasteprene@gmail.com}{https://orcid.org/0000-0002-6862-208X}{}

\authorrunning{R. Brooks et al.} 

\Copyright{R.~Brooks and C.~Hacker and C.~Landi and B.~I.~Mahler and E.~R.~Stephenson} 



\category{} 

\relatedversion{} 

\acknowledgements{Authors are indebted to Asilata Bapat, for her help and generosity.   }

\nolinenumbers 

\EventEditors{John Q. Open and Joan R. Access}
\EventNoEds{2}
\EventLongTitle{42nd Conference on Very Important Topics (CVIT 2016)}
\EventShortTitle{SoCG 2024}
\EventAcronym{SoCG}
\EventYear{2024}
\EventDate{June 11 -- 14, 2024}
\EventLocation{Athens, Greece}
\EventLogo{}
\SeriesVolume{}
\ArticleNo{}

\begin{document}

\maketitle

\begin{abstract}
In multi-parameter persistence, the matching distance is defined as the supremum of weighted bottleneck distances on the barcodes given by the restriction of persistence modules to lines with a positive slope. In the case of finitely presented bi-persistence modules, all the available methods to compute the matching distance are based on restricting the computation to lines through pairs from a finite set of points in the plane.  Some of these points are determined by the filtration data as they are entrance values of critical simplices. However, these critical values alone are not sufficient for the matching distance computation and it is necessary to add so-called {\em switch points}, i.e.~points such that on a line through any of them, the bottleneck matching switches the matched pair. 

This paper is devoted to the algorithmic computation of the set of switch points given a set of critical values. We find conditions under which a candidate switch point is erroneous or superfluous. The obtained conditions are turned into algorithms that have been implemented. With this, we analyze how the size of the set of switch points increases as the number of critical values increases, and how it varies depending on the distribution of critical values. Experiments are carried out on various types of bi-persistence modules.

\end{abstract}

\section{Introduction}
\label{sec:introduction}

The exact computation of the matching distance for multi-parameter persistence modules is an active area of research in computational topology. Besides the theoretical interest \cite{botnan_et_al2022,Bubenik2022,Ethier2023}, one practical reason is that efficient computation of this distance would open the way to the actual application of multi-parameter persistent homology in topological data analysis tasks \cite{biasotti2008multidimensional,Cerri2014,Ethier2014,Wright2020}.

The matching distance is defined  by taking the supremum of the weighted bottleneck distances  on all the (infinitely many) lines with positive slope in the parameter space \cite{Cerri-et-al2013}:

\begin{definition}[Matching Distance]\label{def:matching_distance} The {\em matching distance} between the $n$-persistence modules $M$ and $N$  is defined by
\[ 
\displaystyle d_{match}(M, N) := \displaystyle \sup_{L} \hat m^L \cdot d_B(\dgm M^{L}, \dgm N^{L}) 
\]
\noindent 
where $L\colon u=s\vec m+b$ is any line, with direction $\vec m$ such that $\hat m^L:= \min_i m_i>0$, parametrized by $s\in\RR$,  taken with the standard normalization such that $\max_im_i=1$ and $\sum_ib_i=0$, $M^{L}$ and $N^{L}$ are the restrictions of $M$ and $N$, respectively, to $L$, $\dgm$ denotes the associated persistence diagram, and $d_B$ the bottleneck distance of persistence diagrams. 
\end{definition}

All the available methods for the exact computation of the matching distance are confined to the case $n=2$, i.e.~the matching distance on bi-persistence modules (cf., chronologically, \cite{Kerber-Lesnick-Oudot2018,Bjerkevik2021,bapat2022computing}). This will also be the setting of this paper. Another commonality among these methods is that they make use of a finite set of lines through two kinds of points in the plane. 

On one hand, \cite{Kerber-Lesnick-Oudot2018,Bjerkevik2021,bapat2022computing} use sets of critical grades, such as the sets of entrance grades $C_M$ and $C_N$ of the simplices in the filtration defining the persistence modules $M$ and $N$ respectively, hereafter called {\em critical values}, and their closure with respect to the least upper bound. These points are directly available from the input data. In \cite{Kerber-Lesnick-Oudot2018,Kerber2019} critical values are used to define an initital line arrangement in the dual plane, which is then refined. 

In our work, critical values are used to define an equivalence relation $\sim_{\closure C_M\cup \closure C_N}$ on lines by setting $L\sim_{\closure C_M\cup \closure C_N} L'$ when $L,L'$ partition this set of points in the same way. Interestingly, for any two lines $L,L'$ in the same class, the persistence diagrams $\dgm M^L$ and $\dgm M^{L'}$ (respectively $\dgm N^L$ and $\dgm N^{L'}$) are in bijection, as shown, e.g., in \cite[Thm. 2]{Bapat2022}.  Moreover,  these bijections  extend to a bijection, denoted $\Gamma_{L,L'}$, between the set of matchings between $M^L$ and $N^L$ and the set of matchings between $M^{L'}$ and $N^{L'}$, as shown, e.g., in by \cite{bapat2022computing}. 

Ideally, one would like to use the bijection $\Gamma_{L,L'}$ to show that a line $L$ on which $d_B(M^L,N^L)$ attains a maximum is a line that passes through two points in $\closure C_M\cup \closure C_N$; unfortunately this is not the case, as shown in \cite[Section 2.3]{bapat2022computing}. 
In fact, it is possible that the matched pair of points that obtains the bottleneck distance may ``switch'' within a $\sim_{\closure C_M\cup \closure C_N}$ equivalence class.
Therefore, the equivalence relation $\sim_{\closure C_M\cup \closure C_N}$ must be refined, in order to generate equivalence classes such that there exists at least one matching and matched pair which computes the bottleneck distance for all lines in a given equivalence class.  This requires the addition of a second type of points, hereafter called {\em switch points}: they are points $\omega$ in the projective completion such that the cost of matching some pair $u,v$ equals the cost of matching some other pair $w,x$, with $u,v,w,x\in\closure C_M\cup \closure C_N$ for any line through $\omega$.

As proven in \cite{bapat2022computing}, there exists a finite set of points $\Omega = \Omega(M,N)$ in $\mathbb{P}^2$ with the following property: once the equivalence relation $\sim_{\closure C_M\cup \closure C_N}$ is refined to $\sim_{\closure C_M\cup \closure C_N\cup{ \Omega}}$, at least one matching $\sigma$ and matched pair $u,v\in \closure C_M\cup \closure C_N$ determines the bottleneck distance for all lines in a given equivalence class.  Therefore, the matching distance between $M$ and $N$ may be computed by considering only lines through pairs of points in $\closure C_M\cup \closure C_N\cup{ \Omega}$.

It is important to underline that the formulas derived in \cite{bapat2022computing} for the points of $\Omega$ may produce many unnecessary points. The goal of this paper is to achieve a minimal set of switch points so as to reduce the number of lines needed to compute the matching distance. To this end, we examine the configurations of four points,  not necessarily distinct, used to produce candidate switch points. Each such configuration can produce a set of candidate switch points on the order of $10^3n^4$ where $n=|C_M\cup C_N|$. 
 We identify configurations that cannot produce switch points, as there is no line $L$ with positive slope separating them correctly (cf. \Cref{prop:erroneous-config-3vs1,prop:erroneous-config-2vs2pair}). 
We also identify candidate points that are superfluous based on geometrical considerations (cf. 
\Cref{prop:omega-D,prop:omega-2pv2p,prop:omega-2uv2u}). For the proofs of these results, we refer the reader to \Cref{app:proofs}.

Based on these results, we create functions (given in \Cref{app:pseudo-code}) used to discard configurations of points $u,v,w,x$ that cannot give rise to any switch point and candidate switch points that are superfluous. These functions are implemented in a series of algorithms which, together, generate a complete set of switch points.

We have implemented these algorithms and tested them in various experiments, varying the number and distribution of input critical values. Our tests show that, after the geometric pruning of candidate switch points, the number of unique switch points generated by our algorithms is reduced from the theoretical upper bound by a factor of $3.58\cdot 10^4$. 

{\bf Organization of the paper.} In \Cref{sec:preliminaries} we recall the connection between critical values of bi-persistence modules and points in the persistence diagram of their restriction to a line with positive slope.   In \Cref{sec:switch-points} we review the formulas from \cite{bapat2022computing} to compute candidate switch points. \Cref{sec:results} contains the geometrical results used to refute some candidate switch points as erroneous or superfluous. Finally, in \Cref{sec:experiments} we present the experimental results. All proofs are in \Cref{app:proofs} and pseudocodes are in \Cref{app:pseudo-code}.

\section{Preliminaries}
\label{sec:preliminaries}

We specialize our treatment of persistence to the cases with $1$ or $2$ parameters, referring to them as persistence or bi-persistence, respectively, throughout the paper.

\begin{definition}[Parameter Spaces]
$\RR$ with the usual order $\le$ is the {\em $1D$-parameter space}.  The poset $(\RR^2,\preceq)$  equipped with the following partial order is called  {\em $2D$-parameter space}: for $u=(u_1,u_2),v=(v_1,p_2)\in\RR^2$,  $u\preceq v$ if and only if $u_i\leq v_i$ for  $i=1,2$. 
\end{definition}

The $2D$-parameter space can be sliced by lines with positive slope. A line $L\subset\RR^2$  is a $1D$-parameter space when considered parameterized by $s\in\RR$ as $L: u =\vec m s+b$ where $\vec m\in \RR^2$ and  $b\in \RR^2$. $L$ has positive slope if  each coordinate $m_i$ of $\vec m$ is strictly positive.

\begin{definition}[Persistence Modules] Let $\F$ be a fixed field and $n\in\{1,2\}$. An {\em $n$-persistence module} $M$ over the parameter space $\RR^n$ is an assignment of an $\F$-vector space $M_u$ to each $u\in\RR^n$, and linear maps, called {\em transition} or {\em internal maps}, $i_M(u,v)\colon M_u\to M_v$ to each pair of points $u\preceq v\in\RR^n$, satisfying the following properties:

\begin{itemize}
    \item $i_M(u,u)$ is the identity map for all $u \in\RR^n$.
    \item $i_M(v,w)\circ i_M(u,v)=i_M(u,w)$ for all $u\preceq v\preceq w \in\RR^n$.
\end{itemize}
\end{definition}
  
In this paper, bi-persistence modules will always be assumed to be finitely presented. In particular, for such a module $M$, there exists a finite set  $C_M$ of points in $\RR^2$, called {\em critical values}, for which the ranks of transition maps $i_M(u,v)$ are completely determined by the ranks of transition maps between values in $\closure C_M$, the closure of $C_M$ under least upper bound. For example, one can take $C_M$ to be the set of grades of  generators and relators of $M$. Alternatively,  $C_M$ can be the set of entrance values of cells in a bi-filtration of a cell complex whose homology gives $M$. 

The {\em restriction} of $M$ to a line of positive slope $L\subseteq \RR^2$  is the persistence module $M^L$ that assigns  $M_u$ to each $u\in L$, and whose transition maps $i_{M^L}(u,v)\colon (M^L)_u\to (M^L)_v$ for $u\preceq v\in L$ are the same as in $M$. Once a parameterization $u=\vec ms+b$ of $L$ is fixed, the persistence module $M^L$ is isomorphic to the $1$-parameter persistence module, by abuse of notation still denoted by $M^L$, which
\begin{itemize}
\item assigns to each $s\in\RR$ the vector space $(M^L)_s:=M_u$, and
\item whose transition map between $(M^L)_s$ and $(M^L)_t$ for $s\le t$ is equal to that of $M$ between $M_u$ and $M_v$ with $u=\vec ms+b$ and $v=\vec mt+b$.
\end{itemize}

As $M^L$ and $N^L$ are finitely presented persistence modules, they can be uniquely decomposed as a finite sum of interval modules \cite{Zomordian-Carlsson2005}. A module with underlying support the interval $[b,d)$ can be represented as a point $(b,d)$ in $\RR\times (\RR\cup \{\infty\})$, lying above the diagonal. This encoding permits defining persistence diagrams and the {\em bottleneck distance}  $d_B$ on them \cite{Cohen-Steiner2007}.

\begin{definition}[Persistence Diagram]
If $M\cong \oplus_{j\in J} I[b_j,d_j)^{r_j}$, then the {\em persistence diagram} of $M$, denoted $\dgm M$, is the finite multi-set of points $(b_j,d_j)$ of $\RR\times (\RR\cup\{\infty\})$ with multiplicity $r_j$ for $j\in J$.
\end{definition}

\begin{definition}[Bottleneck Distance] Let $M$, $N$ be two $1$-parameter persistence modules, with persistence diagrams $\dgm M$ and $\dgm N$.  A {\em matching} between $\dgm M $ and $\dgm N$ is a multi-bijection $\sigma: P \rightarrow Q$ between the points of a multi-subset $P$ in $\dgm M$ and a multi-subset $Q$ in $\dgm N$. The {\em cost of a matching $\sigma$}, denoted $c(\sigma)$, is 
\[c(\sigma):=\max \left\{\max_{p\in P}\|p-\sigma(p)\|_\infty,\max_{p\notin P\coprod Q}\frac{p_2-p_1}{2}\right\} .
\]
The {\em bottleneck distance} $d_B$ is then
 \[d_B(\dgm M, \dgm N):= \min_{\substack{\sigma: P\to Q\\ P\subseteq \dgm M, Q\subseteq \dgm N}} c(\sigma). \]
\end{definition}

Briefly, we can say that $c(\sigma)$ is attained by some $\frac{|s-t|}{\delta}$,  with $\delta \in \{1,2\}$ and $s,t$ coordinates of points in $\dgm M\cup \dgm N$.

When the number of parameters is $n= 2$, we can use the bottleneck distance to define the matching distance between persistence modules $M$ and $N$  as in \Cref{def:matching_distance} via restrictions to lines with positive slope. In this case, for convenience, we set ${\rm cost}(\sigma^L):= \hat m^L c(\sigma^L)$ where $\sigma^L$ is a matching between $\dgm M^L$ and $\dgm N^L$ and $\hat m^L=\min_i m_i$.

As shown in \cite{Bapat2022}, it is possible to partition the lines of $\RR^2$ with positive slope into equivalence classes with respect to the relation $\sim_{\overline{C}_M\cup\overline{ C}_N}$ so that, for any two lines $L,L'$ in the same class, the persistence diagrams $\dgm M^L$ and $\dgm M^{L'}$ (respectively $\dgm N^L$ and $\dgm N^{L'}$) are in bijection.  Moreover, by \cite{bapat2022computing}, these bijections  extend to a bijection, denoted $\Gamma_{L,L'}$, between the set of matchings $\sigma^L$ between $M^L$ and $N^L$ and the set of matchings $\sigma^{L'}$ between $M^{L'}$ and $N^{L'}$.
We now review this equivalence relation defined in \cite{Bapat2022}. 

\begin{definition}[Positive Cone]
For every point  $u$ in $\RR^2$, let $S_+(u)$ be the {\em positive cone with vertex $u$}: $S_+(u):=\{v\in\RR^2:u\preceq v\}$. The boundary of the positive cone, $\partial S_+(u)$, decomposes into open faces: 
For $\emptyset\neq A\subseteq\{1,2\}$, define
\[
S_A(u):=\begin{cases}
\{u\} & \textrm { if } A=\{1,2\} \ ,\\
\{(x_1,x_2)\in\RR^2 \  | \ x_1=u_1, x_2>u_2 \}& \textrm{ if } A=\{1\} \ ,\\
\{ (x_1,x_2)\in\RR^2 \  | \ x_1>u_1, x_2=u_2 \} & \textrm{ if } A=\{2\} \ .\\
\end{cases}
\]
\end{definition}

\begin{figure}[H]
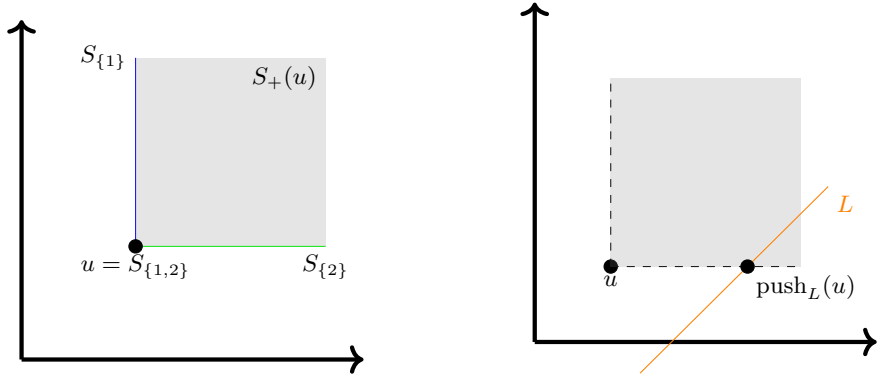

\begin{tabular}{lcr}
\begin{tikzpicture}
\input{Positive_cone.tex} 
\end{tikzpicture} & $\quad$ $\quad$ $\quad$ &
\begin{tikzpicture}
\input{Push.tex}
\end{tikzpicture}
\end{tabular}
\caption{Left: The positive cone $S_+(u)$ of $u\in\RR^2$ and the decomposition of its boundary into $S_{\{1\}}$, $S_{\{2\}}$ and $S_{\{1,2\}}$.
Right: The push of $u$ along the line $L$ when $A^L_u=\{2\}$ ($u$ to the left of $L$).}\label{fig:positive_cone}
\end{figure}

\begin{definition} A line $L$ with positive slope intersects $\partial S_+(u)$ in exactly one point, which we call the {\em push of $u$ onto $L$}, denoted $\mathrm{push}_L(u):=L\cap \partial S_+(u)$. We use the notation $p_L(u)$  for
the parameter value of the push of a point $u$ onto the line $L$. 
\end{definition}

Because of the partition of $\partial S_+(u)$ by its open faces, there is a unique non-empty subset of $\{1,2\}$, denoted $A^L_u$, 
 such that $ \mathrm{push}_L(u)=L\cap  S_{A^L_u}(u)$.   Concretely, in the plane, ${A^L_u}=\{1\}$ means that $u$ is strictly to the right of $L$, ${A^L_u}=\{2\}$ means that $u$ is strictly to the left of $L$, ${A^L_u}=\{1,2 \}$ means that $u$ is on $L$.  More generally, when $\{1\}\subseteq{A^L_u}$, we say $u$ is to the right of $L$, and when $\{2\}\subseteq{A^L_u}$, we say $u$ is to the right of $L$, allowing for the fact that it may be true in either of these cases that $u$ is on $L$.  We say that two lines $L,L'\subseteq \RR^2$ with positive slope have the {\em same reciprocal position} with respect to $u$ if and only if $A^L_u=A^{L'}_u$. 
  Given a non-empty subset $U$ of $\RR^2$, we write $L\sim _{U}L'$ if $L$ and $L'$ have the same reciprocal position with respect to $u$ for all $u \in U$. 
  Note that $\sim_U$ is an equivalence relation on lines with positive slope in $\RR^2$.

As shown in \cite{bapat2022computing}, we need to extend this equivalence relation to include points in the projective completion of the real plane, $\mathbb{P}^2=\RR^2\cup\ell_\infty$,  with  $\ell_{\infty}$ the line at infinity of $\mathbb{P}^2$. Points on the line at infinity are given by homogeneous coordinates $[x_0:x_1:x_2]$ with $x_0=0.$  Note that a line with positive slope in $\RR^2$ will intersect $\ell_{\infty}$ at exactly one point $[0:v_1:v_2]$ with $\vec v= (v_1,v_2)\succ 0$   giving the direction of the line. Given a point $v=[0:v_1:v_2] \in \ell_{\infty}$ with $v_1,v_2>0$, define
\[
S_A(v):=
\begin{cases}
\{v\} & \textrm { if } A=\{1,2\} \ ,\\
\{[0:x_1:x_2]\in \ell_{\infty} \  | \ x_1=v_1, x_2>v_2 \}& \textrm{ if } A=\{1\} \ ,\\
\{[0:x_1:x_2]\in \ell_{\infty} \  | \ x_1>v_1, x_2=v_2 \}& \textrm{ if } A=\{2\} \ .\\
\end{cases}
\]
For every such $v\in \mathbb{P}^2$ and line $L \subset \RR^2$ with positive slope, define $A^L_v$ to be the largest subset of $\{1,2\}$ such that $L \cap S_{A^L_v} \neq \emptyset$. As above, we say that two lines $L,L'$ in $\mathbb{R}^2$ with positive slope are in the {\em same reciprocal position} with respect to $v$ if and only if $A^L_v = A^{L'}_v$.

We then extend the equivalence relation $\sim_U$ form $\RR^2\times \RR^2$ to $\mathbb{P}^2\times \mathbb{P}^2$ as follows: two lines $L$ and $L'$ are equivalent if and only if they are in the same reciprocal position with respect to a finite set $U$ of points in $\mathbb{P}^2$: $
L \sim_U L' \text{ if and only if } A^L_u = A^{L'}_u$ for all $u \in U$.

\section{Explicit formulas for switch points}\label{sec:switch-points}
Let $C_M$ and $C_N$ denote the set of critical values of $M$ and $N$, respectively, as guaranteed by the assumption that $M$ and $N$ are finitely generated, and let $\closure C_M$ and $\closure C_N$ be their closure with respect to the least upper bound.

First, we note that all candidate switch points $\omega$, either proper or at infinity,  can be computed by checking quadruples $u,v,w,x$ (of which at least three must be distinct)  $C_M\cup C_N$  assuming that  $u,v$ and $w,x$ are matched pairs for which, calling $U$ this set of points,
\begin{itemize}
\item there is a line $L$ in a particular equivalence class $E$ via $\sim_U$ on which the cost of matching  $u$ with $v$ equals the cost of matching $w$ with $x$, but
\item the cost of matching $u$ with $v$ {\em does not} equal the cost of matching $w$ with $x$ for all lines in that equivalence class.\end{itemize}

All such lines with this equality in $E$ pass through some point $\omega\in\mathbb{P}^2$, as shown in \Cref{fig:omega3set,fig:omega4set}.  Spanning over all quadruples and all equivalence classes via $\sim_{\closure C_M\cup \closure C_N},$ we obtain the set of candidate switch points $\Omega(M,N)$.
To avoid degenerate or trivial cases, we assume that $u\neq v$ and $w\neq x$.  This is because the matched pair $u,v$ with $u=v$ would indicate a simultaneous birth and death, i.e.~a point on the diagonal within a persistence diagram that yields a null contribution to the matching distance (similarly for $w=x$). However, it can be the case that one of the four points is repeated.

\begin{figure}[H]
	\begin{center}
 		 \begin{minipage}[c]{0.45\textwidth}
		 \begin{center}
			\begin{tikzpicture}
				\input{./figures/OmegaFrom3Points.tex}
			\end{tikzpicture}
			\caption{An example of 3 points $u,w,v=x$ that generate an $\omega$ point with  $m$ denoting the midpoint of $v$ and $x$. 
   }
			\label{fig:omega3set}
		\end{center}
		\end{minipage}\hfill
 		\begin{minipage}[c]{0.45\textwidth}
		\begin{center}
			\begin{tikzpicture}
				\input{./figures/OmegaFrom4Points.tex}
			\end{tikzpicture}
			\caption{An example of 4 points $u,w,v,x$ which generate an $\omega$ point with $m_1$ and $m_2$ denoting the midpoints of  $u$ and $w$, and of $v$ and $x$, respectively. 
   }
			\label{fig:omega4set}
		\end{center}
		\end{minipage}
	\end{center}
\end{figure}

We  also use the following notation:  For $U$ a non-empty set of points in the plane, we  set $A(U)=\{A_u \mid u\in U, \emptyset\neq A_u\subseteq\{1,2\}\}.$  The set $A(U)$ specifies, for each $u \in U$, one or more push directions, and there are $3^m$ possible choices for $A(U)$ if $|U|=m$. Given such a set $A(U)$, we define $E(A(U))$ to be the set of all lines $L$ with positive slope such that $A^L_u = A_u$ for each $u \in U$, i.e. all points of $U$ push onto these lines in the directions specified by $A(U)$.
Note that it is a union of all equivalence classes of $\sim_{\closure C_M\cup \closure C_N}$ in which the points of $U$ push in the specified directions. 

We introduce this notation because switch points are generated based on how the pairs of points $u,v$ and $w,x$ push onto a line. Taking $U=\{u,v,w,x\}$,  this is only a subset of the set $\closure C_M\cup \closure C_N$, implying that there may be many equivalence classes via $\sim_{\closure C_M\cup \closure C_N}$ for which these four points push in the same way for each equivalence class.  So, we create the union of all such equivalence classes  via the set $E(A(u,v,w,x))$.

In \cite[Lemma 4.1]{bapat2022computing}, we proved that, for fixed $U=\{u, v, w, x\} \subseteq \closure C_M\cup \closure C_N$ with $u \neq v$ and $w \neq x$, with at least three distinct points among them,  fixed $\delta, \eta \in \{1,2\}$, and fixed  a choice for $A_u^L, A_v^L, A_x^L, A_w^L$,  if $\omega$ is a switch point for $\sim_U$, then it is determined by one of the following formulas: 
Assuming $p_L(u) \geq p_L(v)$ and $p_L(w)\ge p_L(x)$,
\begin{itemize}
    \item  if $u$, $v$, $w$ lie to the left of $L$ and  $x$ lies strictly to the right, then
    \begin{equation}\label{eq:omega-1}
    \omega=\left[\delta:\delta x_1:\eta(v_2-u_2)+\delta w_2\right].    
   \end{equation}

    \item if $u,v,$ and $w$ lie to the right of lines $L\in E$, and $x$ to the left, then 
    \begin{equation}\label{eq:omega-2}
    \omega=\left[\delta:\eta(v_1-u_1)+\delta w_1:\delta x_2\right].
    \end{equation}

  \item If $u,v$ lie strictly to the left of $L$, $x,w$ strictly to the right,  $x_1\neq w_1$, and $u_2\neq v_2$, then
    \begin{equation}\label{eq:omega-ter}
    \omega=\left[0:\delta(w_1-x_1):\eta(u_2-v_2)\right].
    \end{equation}

    \item If $u,w$ lie strictly to the left of $L$, $x,v$  strictly to the right,  $\delta=\eta$, $x_1\neq v_1$, $u_2\neq w_2$, then
	  \begin{equation}\label{eq:omega-quater}     
     \omega=\left[0: v_1- x_1: u_2 - w_2\right]. 
     \end{equation}

    \item If $u, w$ lie strictly to the left of $L$, $v,x$ lie strictly to the right,  and  $\delta\neq\eta$, then
	\begin{equation}\label{eq:omega-5}    
    \omega=\left[\eta-\delta:\eta v_1 - \delta x_1:\eta u_2 - \delta w_2\right]
     \end{equation}
  \end{itemize}

Assuming $p_L(u) \geq p_L(v)$ and $p_L(w)\leq p_L(x)$,
  \begin{itemize}
      \item if $u$, $v$, $w$ lie to the left of $L$ and  $x$ lies strictly to the right, then
   \begin{equation}\label{eq:omega-3}
	 \omega=\left[\delta:\delta x_1:\eta(u_2-v_2)+\delta w_2\right].
       \end{equation}
     
    \item if $u,v,$ and $w$ lie to the right of lines $L\in E$, and $x$ to the left, then
  \begin{equation}\label{eq:omega-4}
  \omega=\left[\delta:\eta(u_1-v_1)+\delta w_1:\delta x_2\right].
   \end{equation}

      \item If $u, w$ lie strictly to the left of $L$, $v,x$ lie strictly to the right:
	\begin{equation}\label{eq:omega-6}    
    \omega=\left[\eta+\delta:\delta x_1+\eta v_1:\delta w_2+\eta u_2\right].
     \end{equation}
  \end{itemize}
These formulas represent a necessary condition for $\omega$ to be a switch point in each case. More precisely, they are obtained by imposing that 
\[\Delta_L(u,v,w,x;\delta,\eta):=\frac{|p_L(u) - p_L(v)|}{\delta} -\frac{|p_L(w) - p_L(x)|}{\eta}\]
is zero for a line with positive slope $L$, under the chosen conditions for $\delta, \eta$, and $A_u^L, A_v^L, A_x^L, A_w^L$.

\section{Computation of switch points}
\label{sec:results}

To compute all possible switch points, we create algorithms for each of the three cases (listed as cases (2)-(4)
in the proof of \cite[Lemma 4.1]{bapat2022computing}) for which switch points may be generated. We do not include an algorithm for case (1), as this situation does not produce switch points. 
\Cref{alg:3vs1} corresponds to case (2), when three points push in one direction and the fourth point pushes in the other direction;  \Cref{alg:2paired-vs-2paired} corresponds to case (3), when two paired points push in one direction while the other two paired points push in the other direction; finally, \Cref{alg:2unpaired-vs-2unpaired} corresponds to case (4), when two unpaired points push in one direction while the other two unpaired points push in the other direction.

All algorithms share the common assumptions that, among the four points $u,v,w,x$ $u$ is paired to $v$ and $w$ is paired to $x$, and at least 3 of the points $u,v,w,x$ are distinct. The values of both $\delta, \eta$ depend on which of $C_M$ or $C_N$ the points $u,v,w,x$ belong to. Each point can belong to either of the sets, yielding for each set of $u,v,w,x$ multiple combinations of $\delta$ and $\eta$. If $u,v$ are both in $C_M$ (respectively $C_N$), then we need to set $\delta = 2$, and if they are in different sets, then  we need to set $\delta =1$. The same reasoning applies for $w,x$ and $\eta$, yielding up to four possible combinations of $\delta, \eta \in \{1,2\}$. 
 In the algorithms, this is achieved by considering the module a point comes from as a parent of the point. It is possible that $a\in C_M\cap C_N$. In this case, $a$ admits as parents both $M$ and $N$.

Some configurations of points $u,v,w,x$ are not feasible and should be discarded.  In this paper, we address two situations in which computations would lead to erroneous switch points: 
\begin{enumerate}
\item\label{issue-1}  there is no line with positive slope that separates the four points as required;  
\item\label{issue-2} the condition  $\Delta_L=0$ cannot be satisfied by any line with positive slope that separates the four points as required; 

\end{enumerate}

In what follows, we will make sure that the obtained switch points do not present issues as in \Cref{issue-1,issue-2}. We now analyze the three algorithms separately. 
\subsection{Algorithm 3vs1}\label{sec:alg3vs1}

\Cref{alg:3vs1} is used to compute switch points for all possible combinations of four points in $C_M\sqcup C_N$, the corresponding possible combinations of $\delta$ and $\eta$, and the possible choices of $A(u,v,w,x)$ leading to configurations where the point $x$ is strictly on the left (resp. right) of a line in $E(A(u,v,w,x))$ and the points $u,v,w$ on the right (resp. left) of that line, but not strictly. 
This corresponds to the case listed as case (2)
in the proof of \cite[Lemma 4.1]{bapat2022computing}.

To address issue \ref{issue-1}), we introduce the function \textsc{CheckPts} (see \Cref{alg:checkpts}) as a way of discarding some impossible switch points. The correctness of the function \textsc{CheckPts} is guaranteed by the following proposition which identifies when there are no lines in the equivalence class $E(A(u,v,w,x))$: this occurs when $x$ is contained in the convex hull of $u,v,w$, or when $x$ is required to push upward (resp. rightward), but it is to the left and above (resp. to the right and below) of one of the three points.  The following proposition is used to show that the function \textsc{CheckPts} will correctly determine quadruples for which there is no line that separates the four points as required.

\begin{proposition}\label{prop:erroneous-config-3vs1}
Given four not necessarily distinct  points $u,v,w,x$, let  $A_u,A_v,A_w,A_x$ be non-empty subsets of $\{1,2\}$. The following statements hold:

\begin{enumerate}
\item If $2\in A_u\cap A_v\cap A_w$ and $ A_x=\{1\}$, then $E(A(u,v,w,x))$ is empty if and only if there exists $a=(a_1,a_2)$  in the convex hull of $u,v,w$ such that $x_1\le  a_1$ and $x_2\ge a_2$. 
\item If $1\in A_u\cap A_v\cap A_w$ and $ A_x=\{2\}$, then $E(A(u,v,w,x))$   is empty if and only if there exists $a=(a_1,a_2)$ in the convex hull of $u,v,w$, such that $x_1\ge  a_1$ and $x_2\le a_2$.
\item   If $2\in A_u\cap A_v\cap A_w$ and $ A_x=\{1,2\}$,   then  $E(A(u,v,w,x))$   is empty if and only if there exists $a=(a_1,a_2)$  in the convex hull of $u,v,w$ such that $x_1\le  a_1$, $x_2\ge  a_2$, and $x\ne a$.
\item  If $1\in A_u\cap A_v\cap A_w$ and $ A_x=\{1,2\}$,  $E(A(u,v,w,x))$   is empty if and only if there exists $a=(a_1,a_2)$ in the convex hull of $u,v,w$ such that $x_1\ge  a_1$, $x_2\le  a_2$, and $x\ne a$.

\end{enumerate}
\end{proposition}

To address issue 2), we introduce the function \textsc{CheckOmega} (see \Cref{alg:omega-C}) as a way of excluding the switch points which are generated by a quadruple that can be separated appropriately by a line with positive slope, but for which no such line satisfies $\Delta_L=0$. \Cref{prop:omega-C} determines these quadruples; \Cref{prop:omega-D} shows \Cref{alg:checkpts} may be used to partially check the conditions of \Cref{prop:omega-C}.

\begin{proposition}\label{prop:omega-C}
Given three not necessarily distinct points $u,v,w$, and a point $x\ne w,u,v$, let $A_u,A_v,A_w,A_x$ be non-empty subsets of $\{1,2\}$ for which $ E(A(u,v,w,x))$ is non-empty. 

\begin{enumerate}
\item When $2\in A_u\cap A_v\cap A_w$ and $A_x=\{1\}$, letting $\omega$ be given by \cref{eq:omega-1} or  \cref{eq:omega-3} 
, $E(A(u,v,w,x))$ contains some lines $L$ for which  $\Delta_L(u, v, w, x; \delta,\eta)=0$ if and only if
\begin{itemize} 
\item $\omega_2>x_2$ and, moreover,
\item there is no $a=(a_1,a_2)$ in the convex hull of $u,v,w$  with $\omega_1\le a_1$, $\omega_2\ge a_2$, and $a\ne \omega$.
\end{itemize}

\item When $1\in A_u\cap A_v\cap A_w$ and $ A_x=\{2\}$, letting  $\omega$ be given by \cref{eq:omega-2} or \cref{eq:omega-4}, $E(A(u,v,w,x))$ 
contains some lines $L$ for which  $\Delta_L(u, v, w, x; \delta,\eta)=0$ if and only if 
\begin{itemize}
    \item $\omega_1>x_1$ and, moreover,
\item there is no $a=(a_1,a_2)$ in the convex hull of $u,v,w$  with $\omega_1\ge a_1$, $\omega_2\le a_2$, and $a\ne \omega$.
\end{itemize}
\end{enumerate} 
\end{proposition}

\begin{proposition}\label{prop:omega-D}
Given  three not necessarily distinct   points $u,v,w$,  the following statements are equivalent:
\begin{itemize} 
\item there is no $a=(a_1,a_2)$ in the convex hull of $u,v,w$  with $\omega_1\le a_1$, $\omega_2\ge a_2$, and $a\ne \omega$.
\item $\mathrm{CheckPts}(u,v,w,\omega, \mathrm{Below})=\mathrm{True}$, or $\mathrm{CheckPts}(u,v,w,\omega, \mathrm{Below})=\mathrm{False}$ and $\omega\in \partial\mathrm{convhull}(u,v,w)$.
\end{itemize}
Similarly, the following statements are equivalent:
\begin{itemize} 
\item there is no $a=(a_1,a_2)$ in the convex hull of $u,v,w$ with $\omega_1\ge a_1$, $\omega_2\le a_2$, and $a\ne \omega$.
\item $\mathrm{CheckPts}(u,v,w,\omega, \mathrm{Above})=\mathrm{True}$, or $\mathrm{CheckPts}(u,v,w,\omega, \mathrm{Above})=\mathrm{False}$ and $\omega\in \partial\mathrm{convhull}(u,v,w)$.
\end{itemize}
\end{proposition}

We observe that if $\omega$ is on the boundary of the convex hull of $u,v,w$, then the line  through $\omega$ separating $u,v,w$ from $x$ for which $\Delta_L=0$ passes through two of the points $u,v,w$. Thus, this line is eventually generated even if $\omega$ is not added to the set of switch points as it is a line through two critical points. So, this omega point is superfluous and does not need to be added to the set of switch points. \Cref{prop:omega-C,prop:omega-D}, and the last observation justify the call of \Cref{alg:omega-C} by \Cref{alg:3vs1} to reject erroneous or superfluous switch points.

Finally, in \Cref{alg:3vs1}, given a configuration of points $c_1,c_2,c_3,c_4$ extracted from $C_M\sqcup C_N$, such that 3 are distinct, we label them as $x,u,v,w$ so that $x$ and $w$ are distinct, and can then relabel (if necessary) the remaining two points to ensure that $p_L(u)>p_L(v)$.  To determine which switch point is generated, we must also determine the sign of $p_L(x)-p_L(w)$.  The following proposition can be used to determine, from $x$ and $w$, the possible signs of this difference, ensuring the correctness of lines 28--43 of  \Cref{alg:3vs1}.

\begin{proposition}\label{prop:lubparametercheck}
    Let $x,w$ be two distinct points in $C_M\cup C_N$. For every line $L$  with positive slope which separates $x$ and $w$, it holds that: 
    \begin{itemize}
    \item If $\lub(w,x)=x$, then $p_L(x)\geq p_L(w)$.
    \item If $\lub(w,x)=w$, then $p_L(x)\leq p_L(w)$.
    \item If $\lub(w,x)$ is neither $x$ nor $w$, then $p_L(x)\geq p_L(w)$ for all lines $L$ which separate $w$ and $\lub(w,x)$, and $p_{L'}(x)\leq p_{L'}(w)$ for all lines $L'$ which separate $x$ and $\lub(w,x)$.
    \end{itemize}
\end{proposition}

\subsection{Algorithm 2paired-vs-2paired}\label{sec:alg-2paired-vs-2paired}

\Cref{alg:2paired-vs-2paired} is used to compute switch points for all possible combinations of four points in $C_M\sqcup C_N$,  the corresponding possible combinations of $\delta$ and $\eta$, and the possible choices of $A(u,v,w,x)$ leading to configurations where two paired points lie strictly on the left of a line in $E(A(u,v,w,x))$ and the other two paired points lie strictly on the right of that line.  This corresponds to the case listed as case (3) in the proof of \cite[Lemma 4.1]{bapat2022computing}. 

In case (3) of \cite[Lemma 4.1]{bapat2022computing},  we saw that if $ A_u= A_v=\{2\}$ and $A_x=A_w=\{1\}$, then no switch point is generated if $x_1=w_1$ or $u_2=v_2$. Moreover, 
since the pairs must lie strictly on one side of the line, it is not possible that one point may simultaneously be in two pairs. Since paired points must be distinct (otherwise, the cost of matching that pair would be zero), it must be the case that all 4 points are distinct.

If a quadruple of distinct points generates a switch point $\omega$ with this configuration, it is given by \Cref{eq:omega-ter}. Note that, given four distinct points that are separated by a line $L$ with positive slope so that two lie strictly on one side of the line and the other two lie strictly on the other side, there is exactly one way to label these points as $x,w,u,v$ so that $x,w$ push up to $L$ and $p_L(w)>p_L(x)$, and $u,v$ push right to $L$ and $p_L(u)>p_L(v)$.
With this labeling, the slope that is generated by \Cref{eq:omega-ter} is by construction positive.

Next, we introduce the function \textsc{CheckPts2} (see \Cref{alg:checkpts-2v2paired}) to discard sets of points which cannot be separated by a line with positive slope as required (issue \ref{issue-1}).  The correctness of this function is guaranteed by \Cref{prop:erroneous-config-2vs2pair}. More precisely, the function \textsc{CheckPts2} uses condition 4 from \Cref{prop:erroneous-config-2vs2pair} to check for erroneous quadruples of points.

\begin{proposition}\label{prop:erroneous-config-2vs2pair}
Let $u,v,w,x$ be four points, with $ A_u= A_v=\{2\}$, $A_x=A_w=\{1\}$. Set 
\[Q=\{(p_1,p_2)\in\RR^2|p_1
     \leq a_1 \textrm{ and } p_2\geq a_2\textrm{ for some } a\in \overline{uv}\} \textrm{  and  }\]
    \[R=\{(p_1,p_2)\in\RR^2|p_1
     \geq b_1 \textrm{ and } p_2\leq b_2\textrm{ for some } b\in \overline{xw}\}.\]
     The following statements are equivalent:

     \begin{enumerate}
         \item[(i)] $R\cap Q\neq\emptyset$,
         \item[(ii)] the set of lines $E(A(u,v,w,x))$ is empty,
         \item[(iii)] there are a point $a=(a_1,a_2)$  on the line segment $\overline{uv}$ and a point $b=(b_1,b_2)$ on the line segment $\overline{xw}$ such that $(b_1\le  a_1)\wedge (b_2\ge a_2)$,
         \item[(iv)] Either $x$ or $w\in Q$, or $u$ or $v\in R.$
     \end{enumerate}
\end{proposition}

The following proposition determines if it is possible to split  four points appropriately with a line with slope determined by the $\omega$ given in \Cref{eq:omega-ter} (issue \ref{issue-2}).  This will happen exactly when the $y$-intercepts of the lines through $u,v,w,x$ are ordered appropriately. \Cref{prop:omega-2pv2p} ensures the correctness of line 31 in \Cref{alg:2paired-vs-2paired}.

\begin{proposition}\label{prop:omega-2pv2p}
 Given four distinct points $u,v,w,x$, let $A_u=A_v=\{2\}$ and $A_w=A_x=\{1\}$, and assume that the set $ E(A(u,v,w,x))$ is non-empty. Without loss of generality, assume that $p_L(u)>p_L(v)$ and $p_L(w)>p_L(x)$. 
 Then, for every $\eta,\delta\in\{1,2\}$, letting $m=\frac{\delta(w_1-x_1)}{\eta(u_2-v_2)}$ be the slope identified by $\omega$ as given in \cref{eq:omega-ter}, the equivalence class $E(A(u,v,w,x))$ contains some lines $L$ for which  $\Delta_L(u, v, w, x; \delta,\eta)=0$ if and only if 
 \begin{equation}\label{eq:omega-2pv2p}
     \max\{x_2-mx_1,w_2-mw_1\}<\min\{u_2-m u_1,v_2-mv_1\}
 \end{equation}
\end{proposition}

\subsection{Algorithm 2unpaired-vs-2unpaired}\label{sec:alg-2unpaired-vs-2unpaired}

\Cref{alg:2unpaired-vs-2unpaired} is used to compute switch points for all possible combinations of four points in $C_M\sqcup C_N$, with at least three of them distinct,  the corresponding possible combinations of $\delta$ and $\eta$, and the possible choices of $A(u,v,w,x)$ leading to configurations where a line in $E(A(u,v,w,x))$ splits both pairs; i.e., one point of a pair is strictly on the left, and one point is strictly on the right. This corresponds to the case listed as case (4)
in the proof of \cite[Lemma 4.1]{bapat2022computing}. We may assume without loss of generality  that $u$ and $w$ lie to the left of the line, and that $v$ and $x$ lie to its right. Depending on $L$,  both $p_L(u)\geq p_L(v)$ and $p_L(u)\leq p_L(v)$  can occur (similarly for $x,w$). To decide which one occurs, we apply \Cref{prop:lubparametercheck} in lines 28 and 42  of \Cref{alg:2unpaired-vs-2unpaired}.

To address issue \ref{issue-1}) we use again the function \textsc{CheckPts2} (\Cref{alg:checkpts-2v2paired}) simply by changing the roles of $u,v,w,x$; calling $\textsc{CheckPts2}(x,v,w,u)$ instead of $\textsc{CheckPts2}(x,w,u,v)$. If the configuration is feasible, we compute the candidate switch points with the following formulas:
\begin{itemize}
\item If  $p_L(u)\geq p_L(v)$ and $p_L(w)\geq p_L(x)$, or $p_L(u)\leq p_L(v)$ and $p_L(w)\leq p_L(x)$, and moreover
\begin{itemize}
    \item $\delta=\eta$, and $x_1= v_1$ or $u_2= w_2$, then no switch point is generated;
    \item $\delta=\eta$,  $x_1\neq v_1$, and $u_2\neq w_2$, then $\omega$ is given by \Cref{eq:omega-quater}; 
    \item  $\delta\ne \eta$, then $\omega$ is given by \Cref{eq:omega-5}.
\end{itemize}
\item If  $p_L(u)\geq p_L(v)$ and $p_L(w)\leq p_L(x)$, or  $p_L(u)\leq p_L(v)$ and $p_L(w)\geq p_L(x)$, then $\omega$ is given by \Cref{eq:omega-6}.
\end{itemize}

 In the case when $\omega$ is a point at infinity, issue \ref{issue-2}) can be addressed as in \Cref{prop:omega-2pv2p} with the roles of $w$ and $v$ switched.  The only assumption from \Cref{prop:omega-2pv2p} that is not guaranteed is $p_L(u)>p_L(v)$ and $p_L(w)>p_L(x)$, which, in the proof, ensures that any line passing through $\omega$ has positive slope.  So, in \Cref{alg:2unpaired-vs-2unpaired}, we discard any $\omega$ points which generate lines with non-positive slope, and then check a modified version of \Cref{eq:omega-2pv2p} (line 32). When $\omega$ is a proper point, we address issue \ref{issue-2}) by introducing the function \textsc{CheckOmega2} (see \Cref{alg:checkomega3}).  The correctness of \Cref{alg:checkomega3} is given by \Cref{prop:omega-2uv2u}.

\begin{proposition}\label{prop:omega-2uv2u}
 Given four points $u,v,w,x$, let $A_u=A_w=\{2\}$ and $A_v=A_x=\{1\}$ 
and assume that $ E(A(u,v,w,x))$ is non-empty. Let $\omega$ be the point generated by $x,v,u,w$ by \cref{eq:omega-5} or \cref{eq:omega-6}. Partition $\RR^2$ into quadrants with respect to $\omega$ in the following way:
\begin{align*}
    Q_1&=\{(p_1,p_2)\in\RR^2|p_1>\omega_1 \textrm{ and }p_2>\omega_2 \},\quad
    Q_2&=\{(p_1,p_2)\in\RR^2|p_1\leq\omega_1 \textrm{ and }p_2\geq\omega_2 \}\\
   Q_3&=\{(p_1,p_2)\in\RR^2|p_1<\omega_1 \textrm{ and }p_2<\omega_2 \},\quad 
    Q_4&=\{(p_1,p_2)\in\RR^2|p_1\geq\omega_1 \textrm{ and }p_2\leq\omega_2 \}
\end{align*}

\begin{enumerate}
    \item If either $x$ or $v$ $\in Q_2$, or $u$ or $w$ $\in Q_4$, there is no line through $\omega$ in $E(A(u,v,w,x))$.
    \item If  $x,v\in Q_4$, $u,w\in Q_2$, and $\omega\notin\{x,v,u,w\}$, all lines through $\omega$ with positive slope are in $E(A(u,v,w,x))$.
    \item Suppose that $x,v\notin Q_2$, $u,w\notin Q_4$, and $\{u,v,x,w\}\cap(Q_1\cup Q_3)\neq\emptyset$. For $c\in\{u,v,x,w\}\cap(Q_1\cup Q_3)$, let $0\le m_c\le \infty$ be the slope of the line through $\omega$ and $c$, and define
    \[
    \begin{split}
          \mathrm{sign}_1(c)=\begin{cases} +1 & \textrm{ if } c\in Q_1\\-1 &  \textrm{ if } c\in Q_3
          \end{cases}
          &   \qquad \textrm{ and }   \qquad     \mathrm{sign}_2(c)=\begin{cases} +1 & \textrm{ if } c\in \{x,v\}\\-1 &  \textrm{ if } c\in \{u,w\}.
          \end{cases}
    \end{split}
    \]
    
    Then there is a line with positive slope through $\omega$ in $E(A(u,v,w,x))$ if and only if there exists $0<m<\infty$ such that, for every $c\in \{u,v,x,w\}\cap(Q_1\cup Q_3)$, \[\mathrm{sign}_1(c)\mathrm{sign}_2(c)m_c < \mathrm{sign}_1(c)\mathrm{sign}_2(c)m.\] 
    
\end{enumerate}

\end{proposition}

\section{Experiments}\label{sec:experiments}

\begin{figure}[h!]
    \includegraphics[width=0.6\textwidth]{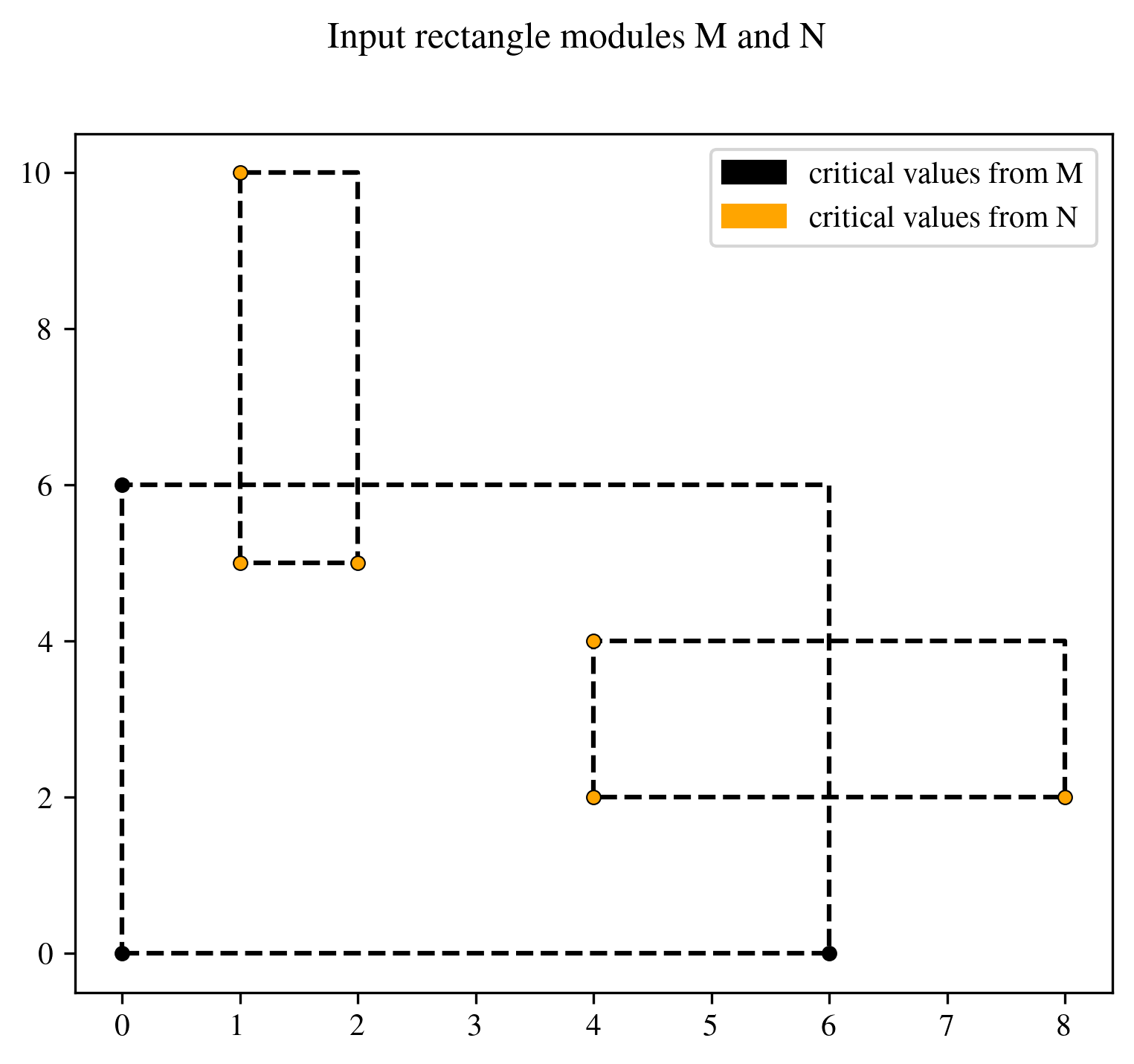}
    \includegraphics[width=0.4\textwidth]{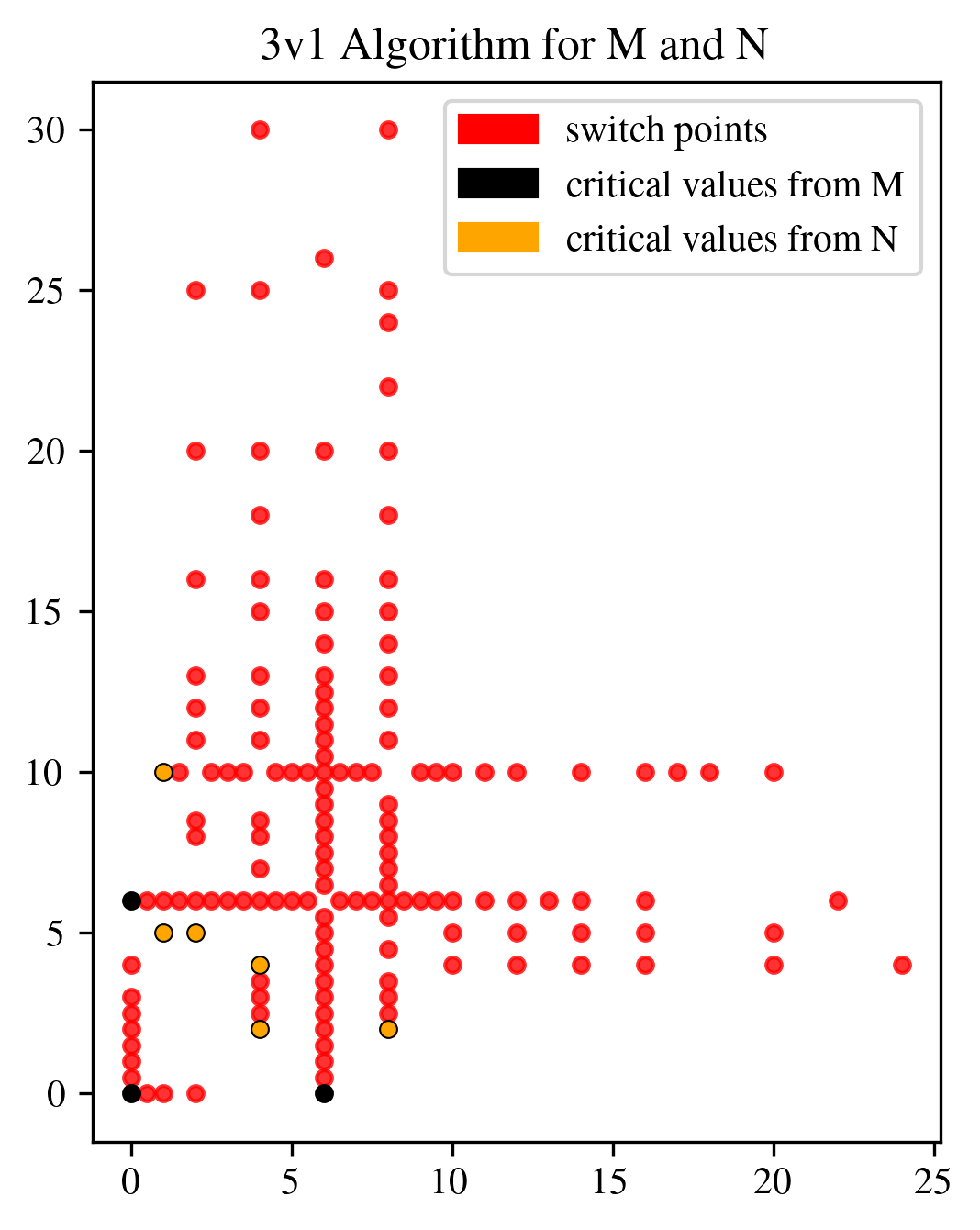}
    \includegraphics[width=0.6\textwidth]{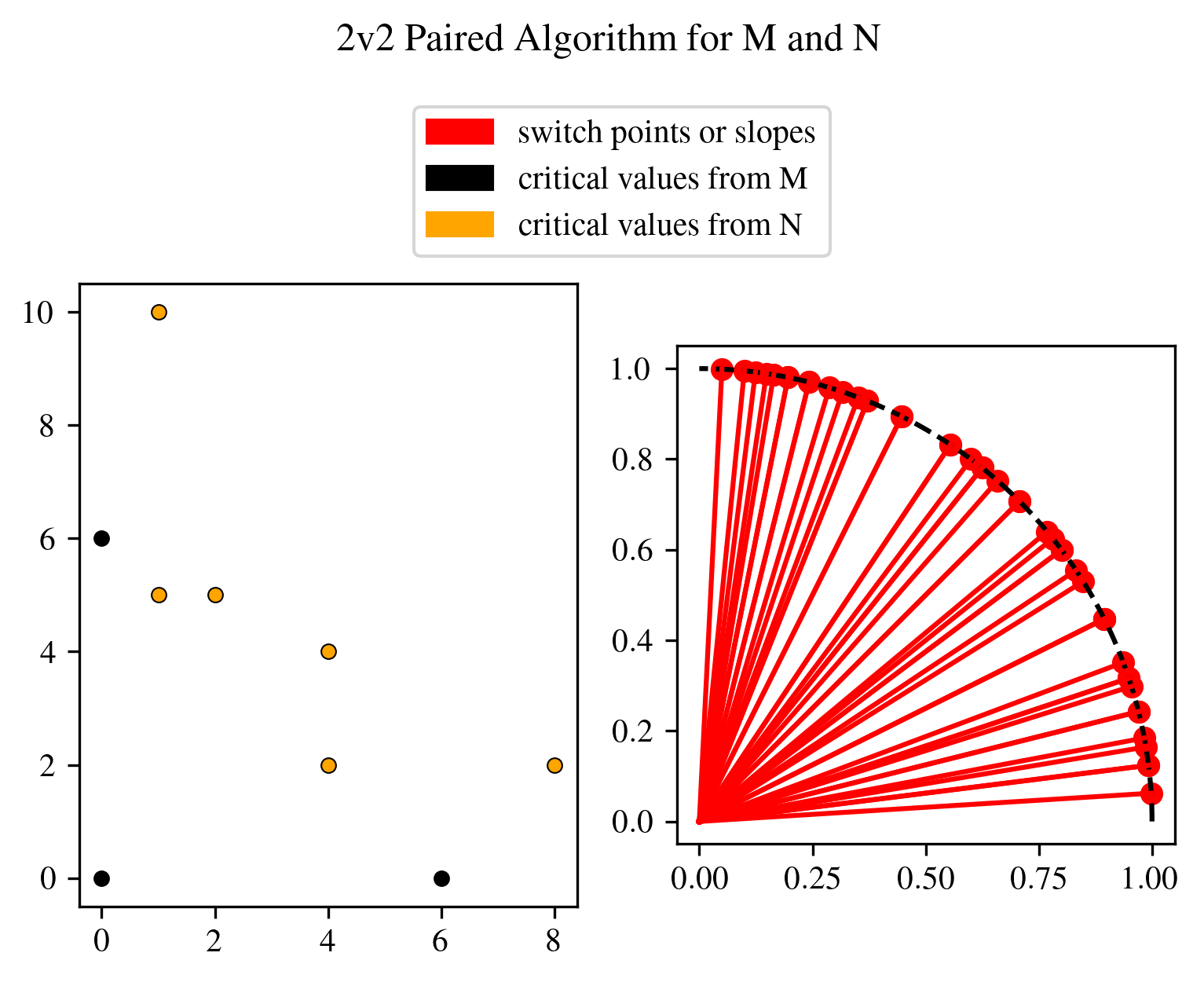}
    \includegraphics[width=0.6\textwidth]{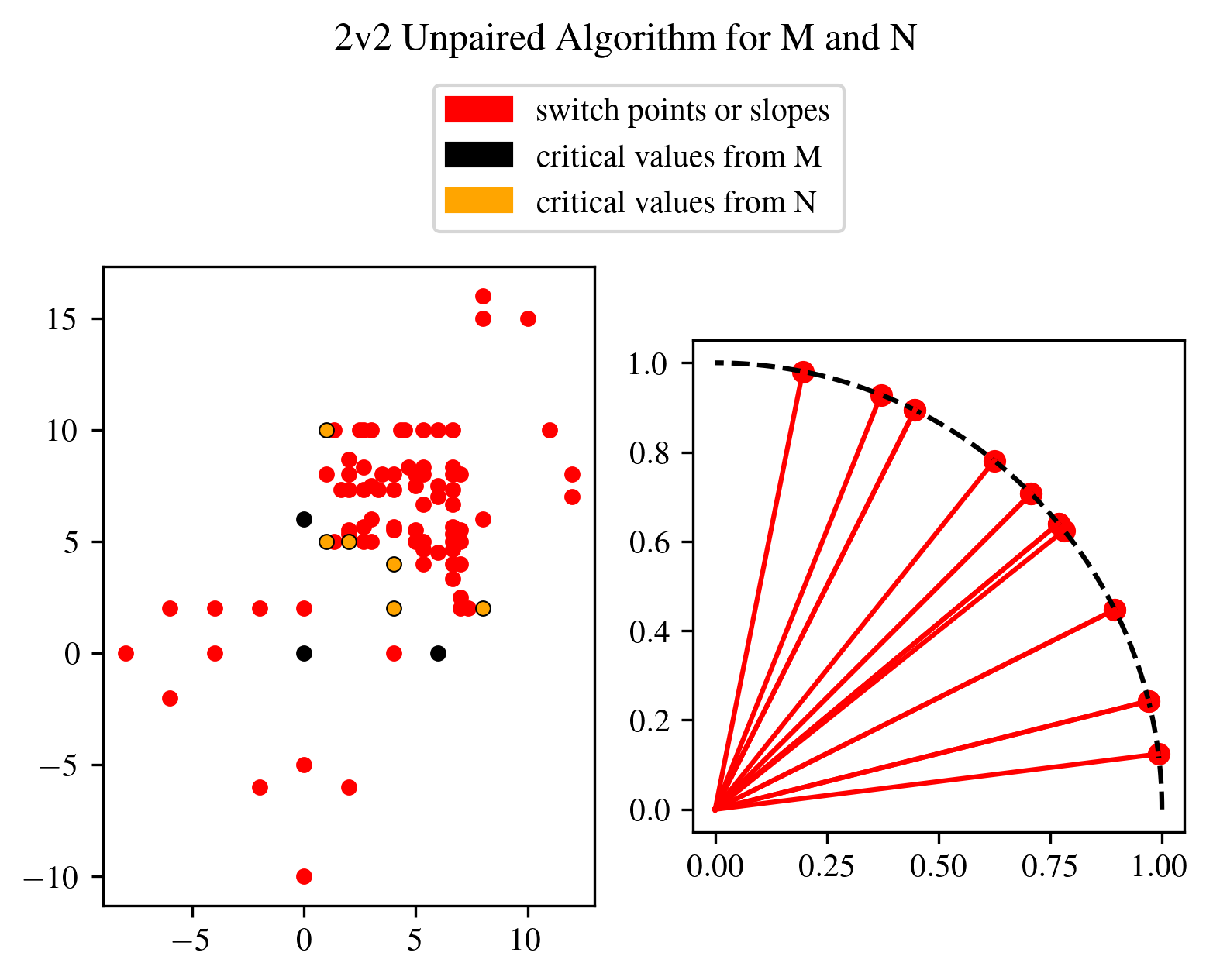}
    \caption{
    For modules $M$ and $N$ given in the top left, \Cref{alg:3vs1} computed 914 switch points, and deduplicated them to 142. \Cref{alg:2paired-vs-2paired} computed 133 slopes (with duplicates), and 44 without duplicates. Note that all of the outputs are points at infinity. \Cref{alg:2unpaired-vs-2unpaired} computed 181 switch points and points at infinity combined (with duplicates) and without duplicates 95.}
    \label{fig:MandNandOmegaPts}
\end{figure}

First, \Cref{fig:MandNandOmegaPts} shows the outputs of \Cref{alg:3vs1,alg:2paired-vs-2paired,alg:2unpaired-vs-2unpaired} for an example pair of modules.

Next, we consider the theoretical upper bound of candidate switch points. Given two sets of critical values $C_M$ and $C_N$ with $n=|C_M\cup C_N|$, there are $n(n-1)(n-2)(n-3)$ possible subsets of four distinct points $c_1,c_2,c_3,c_3$, and  $n^2(n-1)(n-2)$ possible subsets of four points with one repetition, summing up to $n(n-1)(n-2)(2n-3)$ choices. For each such choice, there are 4 choices of labeling one of them as $x$ or $u$, and 3 as $w$ or $v$, yielding 144 choices for each set $c_1,c_2,c_3,c_3$.  As \Cref{alg:3vs1} produces at most 4 switch points, \Cref{alg:2paired-vs-2paired} at most 1, and \Cref{alg:2unpaired-vs-2unpaired} at most 2, we deduce that in the worst case the number of switch points is on the order of  $10^3n(n-1)(n-2)(2n-3)$.

\begin{figure}[h!]

\begin{center}
\includegraphics[width=0.4\textwidth]{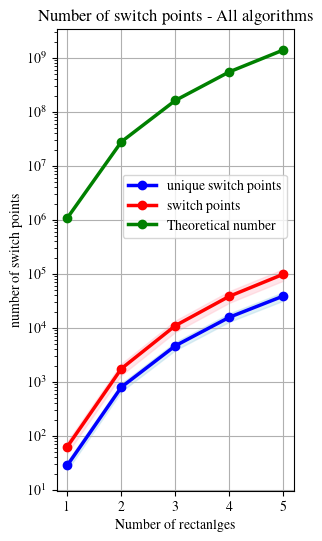}
    \caption{We compute the number of switch points generated for pairs of random rectangle decomposable bi-persistence modules with the same number of rectangles ($x$-axis). We record the number of switch points, unique switch points and the theoretical upper bound. The average and standard error of the mean are computed over 5 runs. 
  }\label{fig:experimentalomeganumber}
\end{center}
\end{figure}

In \Cref{fig:experimentalomeganumber}, we show how the computed number of switch points increases with the number of critical values.  We see that the number of unique switch points is smaller than the total number of computed switch points; on average, the ratio of unique points to computed points is $0.55$. However, the number of computed switch points is much smaller than the theoretical bound; on average, the ratio of total computed points to the theoretical bound is $6.57\cdot 10^{-5}$.  
This is because the \Cref{alg:3vs1,alg:2paired-vs-2paired,alg:2unpaired-vs-2unpaired} discard repeated quadruples which only differ by relabeling, and due to functions \textsc{CheckPoints}, \textsc{CheckPoints2}, \textsc{CheckOmega}, and \textsc{CheckOmega2}.

\begin{figure}[h!]

    \includegraphics[width=0.5\textwidth]{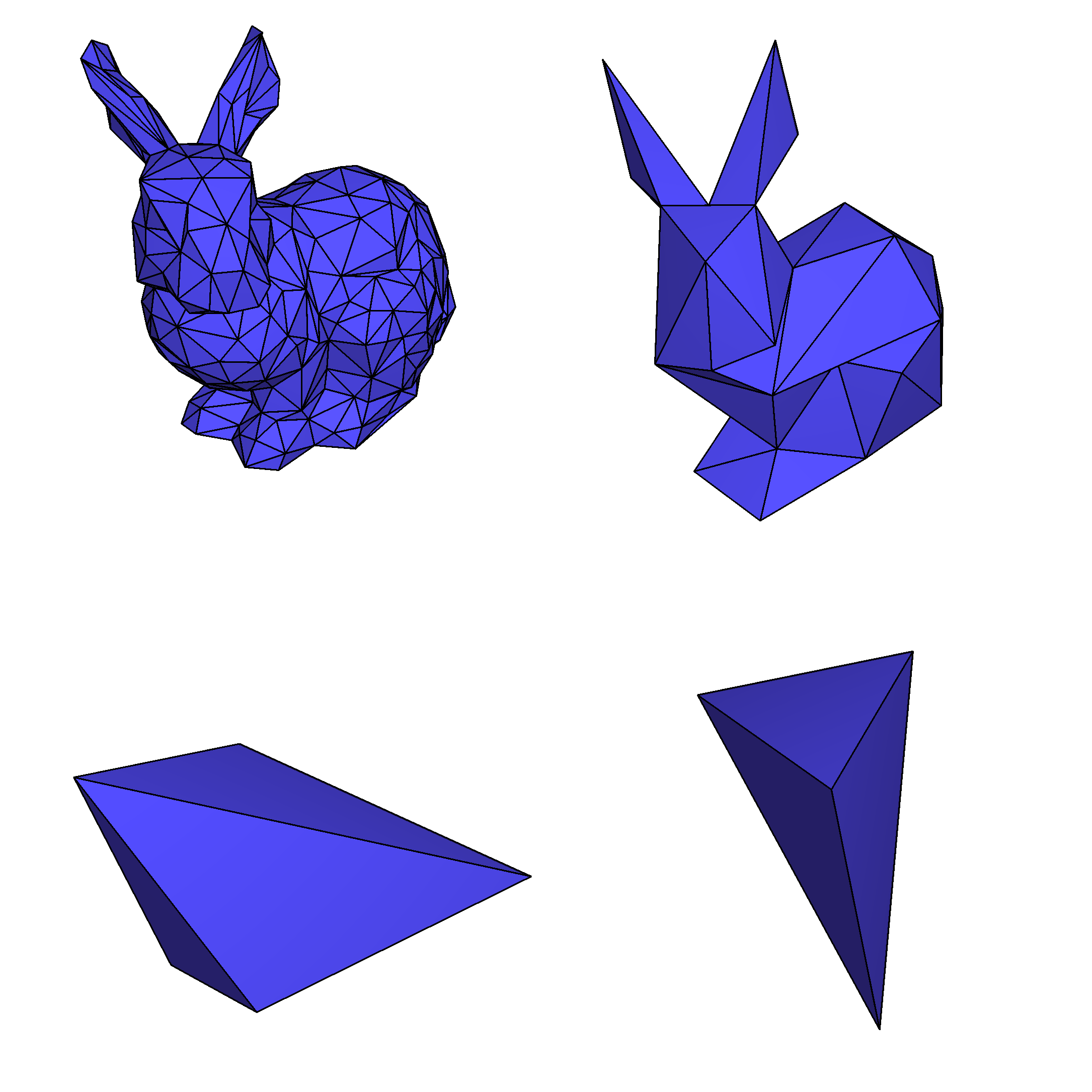}
  \includegraphics[width=0.5\textwidth]{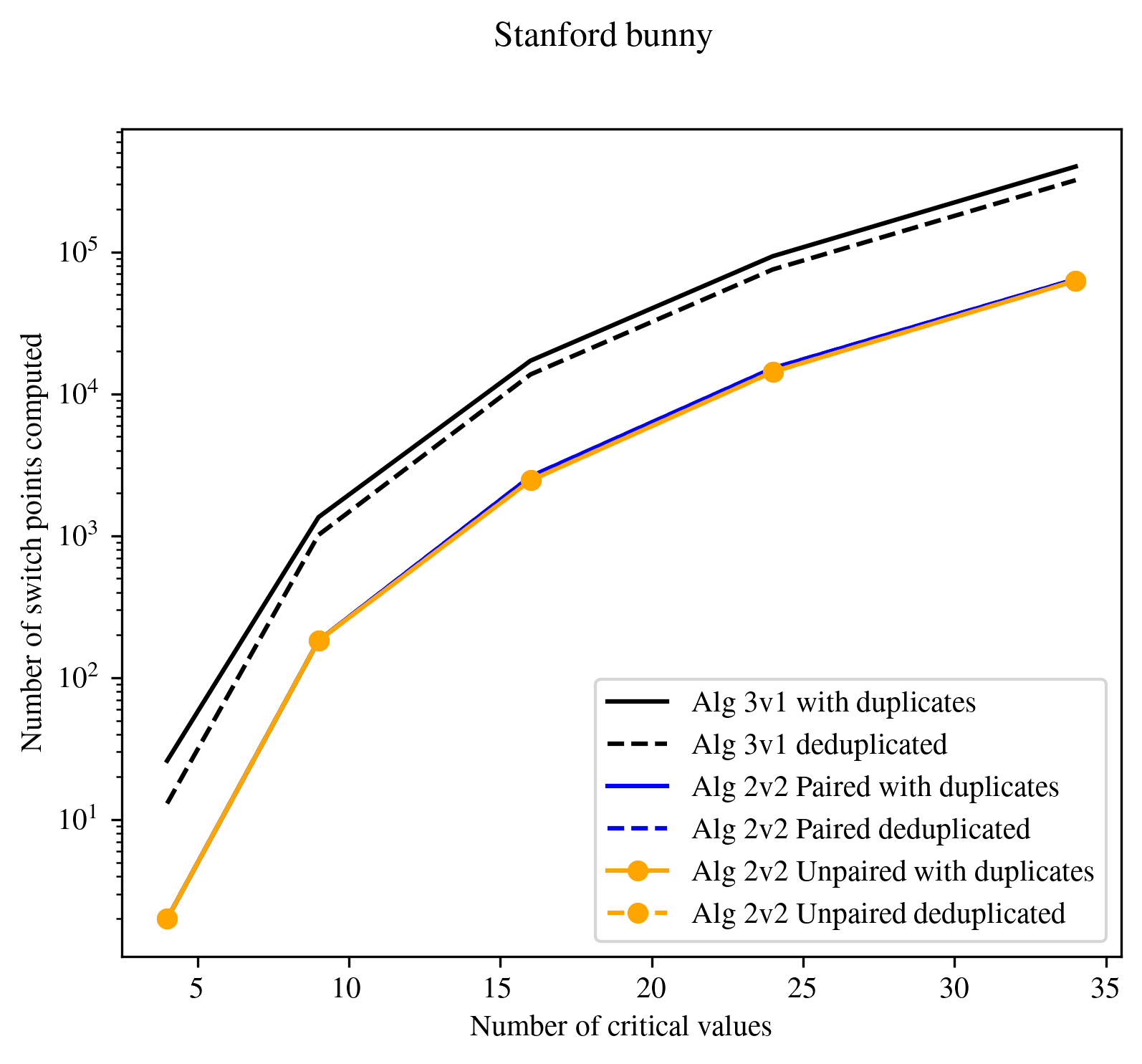}
  \includegraphics[width=0.5\textwidth]{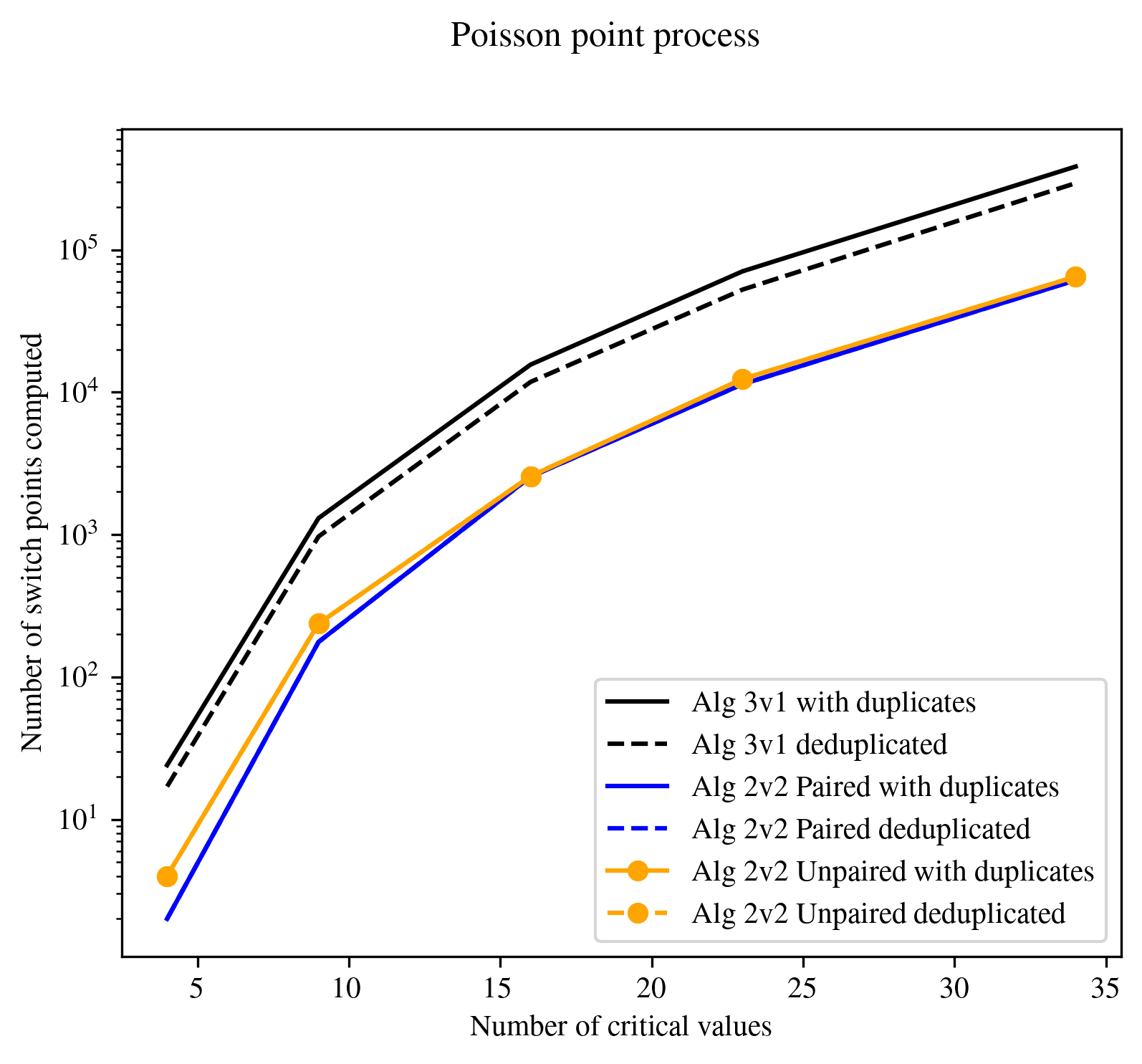}
  \includegraphics[width=0.5\textwidth]{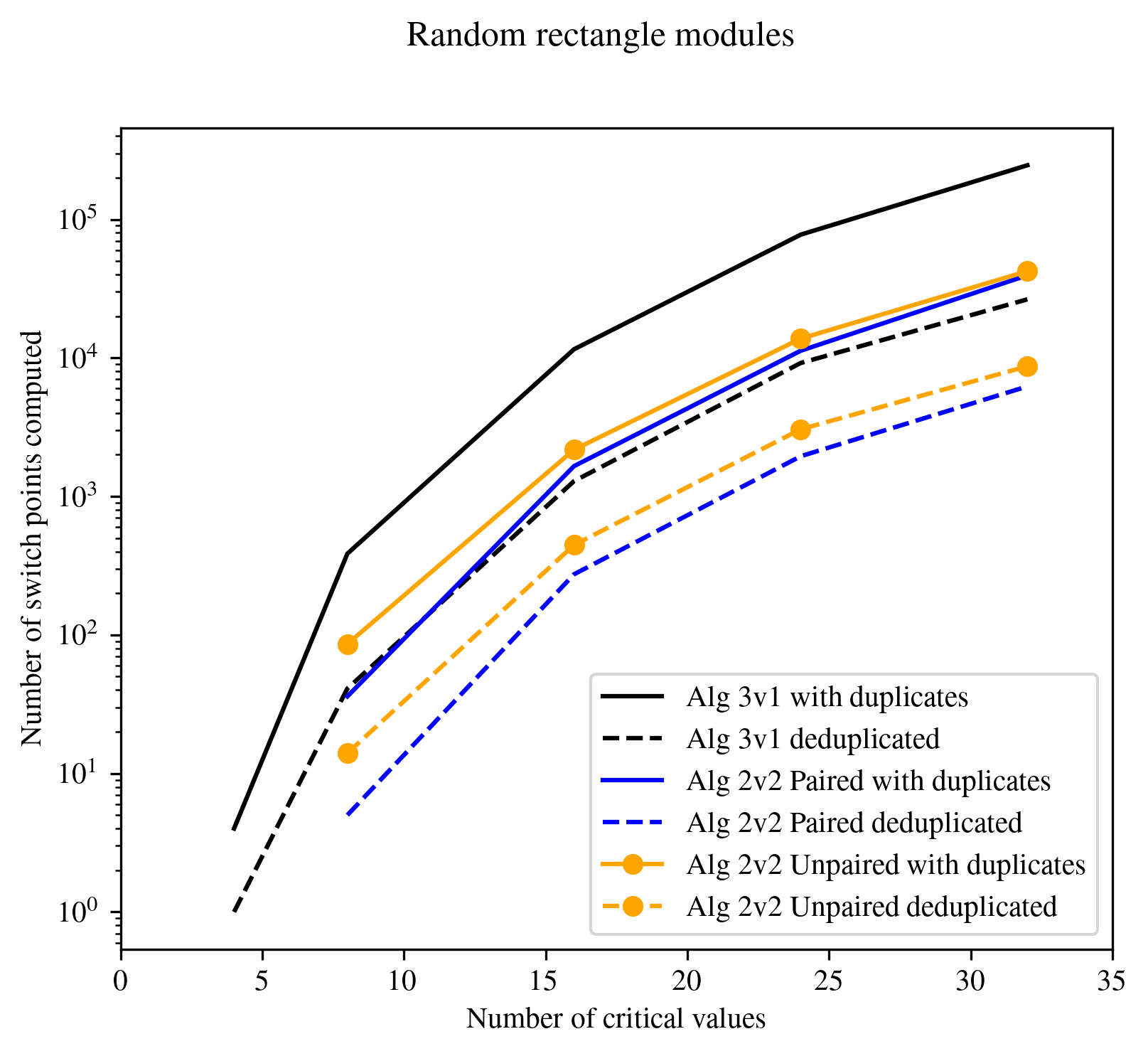}
  
    \caption{Here we compare the number of switch points generated from differently organised input critical values: namely, the vertices of the decimated Stanford Bunny projected to the $xy$-plane, the output of a Poisson point process with blue noise, and randomly generated rectangle decomposable bi-persistence modules.
  }
    \label{fig:bunny}
\end{figure}

In \Cref{fig:bunny} we compare the outputs for three data sets:  the output of a Poisson point process with blue noise, data with real-world attributes (noisy, less regular), and data from a rectangle module. The results show that on data with fewer symmetry constraints than in the rectangle case, the number of duplicate switch points almost vanishes. 

\section{Conclusions}
\label{sec:conclusions}

Following the theory of \cite{bapat2022computing}, we have provided algorithms to generate a set of candidate switch points necessary to compute the matching distance between two finitely generated bi-persistence modules.  Further, we have provided conditions and algorithms for excluding erroneous and superfluous candidate switch points, thereby improving the computation of the matching distance (via computing the bottleneck distance of restrictions to lines through pairs from the union of the closure of critical points and switch points). To our knowledge, while switch points are used in previous work on the computation of the matching distance, the issue of making explicit a minimal set of switch points had not been addressed before. 

We have also presented experiments on different types of data which show that our implementation can significantly lower the number of computed switch points from the theoretical bound, showing a slightly different behavior with respect to the number of duplicates.

As any algorithm for the exact computation of the matching distance runs in polynomial time on the number of points used to generate the lines for the restriction, this work has the potential to greatly reduce the overall computation time.

Our results fill an important gap in the computation of the matching distance. However, some questions remain to be answered. First, while we have implemented code to generate the necessary switch points, we have yet to integrate this into a larger algorithm that directly computes the matching distance between two modules.  

Second, while we have been able to discard erroneous and superfluous candidate switch points caused by issues \ref{issue-1} and \ref{issue-2}, there are still candidate switch points in the set generated by \Cref{alg:3vs1,alg:2paired-vs-2paired,alg:2unpaired-vs-2unpaired} which can be discarded, because of the following issue.  When a candidate switch point $\omega$ generated by $x,u,v,w$ lies outside the support of the indecomposable summands associated with $x,u,v,w$, it may be the case that obtained bars (generated by $x,u,v,w$) on a line through $\omega$ do not actually ``switch'' matching.  The switch point is generated because the cost of matching the projections of the critical values onto the line is equal, but since $\omega$ is outside the support, the cost of matching the bars may not be equal.  

For example, in \Cref{fig:MandNandOmegaPts} (\Cref{alg:3vs1}), there are many switch points outside the support of the critical values with $x$-coordinate $6$.  However, the bar that will be generated by the critical values from $M$ will not extend past the support of that rectangle. So, it is likely that such candidate switch points may also be discarded. However, it is not possible to discard such switch points with the assumptions we are making, as this would require knowledge on the support of the persistence modules $M$ and $N$ and their decomposition.



\bibliography{switch_points_arxiv}

\appendix

\section{Proofs}\label{app:proofs}
\addcontentsline{toc}{section}{Appendix}

\input

\begin{proof}[Proof of \Cref{prop:erroneous-config-3vs1}]
Let us start by proving the first statement. The second one can be proved similarly. We refer to \Cref{fig:erroneous3vs1} for an illustration.

\begin{figure}[htb]
    \begin{center}
    \begin{subfigure}{0.45\textwidth}
                \begin{center}
			\begin{tikzpicture}
				\input{./figures/xPushUpConvexHullExample1.tex}
			\end{tikzpicture}
                \end{center}
    \caption{Case $\bar a\in\mathrm{convhull}(u,v,w)$: For any $x$ in the upper left quadrant $Q$, there is an $a$ in the convex hull of $u,v,w$ for which $x_1\le  a_1$, $x_2\ge a_2$ and therefore the set of lines with positive slope which separate $x$ from $u,v,w$ is empty.  }
    \label{fig:convexhullexample1}
    \end{subfigure}
 \hfill
        \begin{subfigure}{0.45\textwidth}
        \begin{center}
        		\begin{tikzpicture}
				\input{./figures/xPushUpConvexHullExample2.tex}
			\end{tikzpicture}
        \end{center}
        \caption{Case: $\bar a\notin \mathrm{convhull}(u,v,w)$: For any $x$ in the upper left quadrant $Q$ which is not in the yellow region, there is an $a$ in the convex hull of $u,v,w$ for which $x_1\le  a_1$, $x_2\ge a_2$ and therefore the set of lines with positive slope which separate $x$ from $u,v,w$ is empty.  If $x$ is in the yellow region, the line through $z_r$ and $z_l$ has positive slope and separates $x$ from $u,v,w$.}
	\label{fig:convexhullexample2}
  \end{subfigure}
\end{center}
 \caption{ Examples of configurations of points for \Cref{prop:erroneous-config-3vs1}, case 1.
 In both figures, we may use vertical and horizontal lines through $\bar a$ to partition the plane into four quadrants. For points $x$ to the right of the vertical line through $\bar a$, we may rotate that line slightly clockwise around $\bar a$ to obtain a line with positive slope separating $x$ from $u,v,w.$ For points $x$ below the horizontal line through $\bar a$, we may rotate that line slightly counterclockwise around $\bar a$ to obtain a line with positive slope separating $x$ from $u,v,w$. For points $x$ such that  $x_1\le \bar a_1$ and $x_2\ge \bar a_1$, the cases $\bar a\in \mathrm{convhull}(u,v,w)$ and $\bar a\notin \mathrm{convhull}(u,v,w)$ must be studied separately.}
\label{fig:erroneous3vs1}
\end{figure}

Let us assume there is a point $a=(a_1,a_2)$  in the convex hull of $u,v,w$  such that $x_1\le  a_1$ and $x_2\ge a_2$. As $a$ is in the convex hull of $u,v,w$, if there were a  line $L:y_2=my_1+q$ with slope $m>0$ strictly separating $x$ from $u,v,w$, then $L$ would also separate $x$ from $a$. So, $a_2\ge ma_1+q$ while $x_2< mx_1+q$. Now, the assumptions $x_1\le a_1$ and $x_2\ge a_2$ imply $x_2\ge mx_1+q$, i.e. that $x$ lies above $L$, yielding a contradiction. Therefore, the existence of such point $a$ implies that the subset of the lines of positive slope that do not contain $x$ in  $E(A(u,v,w,x))$ is empty.

Vice versa, let us assume that there is no point $a = (a_1, a_2)$ in the convex hull of $u$, $v$, $w$ such that $x_1 \le a_1$ and $x_2 \ge a_2$. Let $\bar a_1\in \RR$ be the maximum abscissa of points $a$ in the convex hull of $u,v,w$, and $\bar a_2\in \RR$ their minimum ordinate. Set $\bar a=(\bar a_1,\bar a_2)\in \RR^2$. 

In the case when $\bar a$ belongs to the convex hull of $u,v,w$, by assumption we have $x_1>\bar a_1$ or $x_2<\bar a_2$. If $x_1>\bar a_1$, let us consider the vertical line $L'$ through $\bar a$ so that $x$ is strictly to the right of $L'$. 
No point of $L'$ below $\bar a$ and no point to the right of $L'$ belong to the convex hull of $u,v,w$. 
Thus,  slightly rotating clockwise the line $L'$ about $\bar a$, we obtain a valid line $L\in E(A(u,v,w,x))$. Similarly, if $x_2<\bar a_2$, let us consider the horizontal line $L''$ through $\bar a$ so that $x$ is strictly below $L''$. 
No point of $L''$ to the right of $\bar a$ and no point below $L''$ belongs to the convex hull of $u,v,w$. 
Thus,  slightly rotating counter-clockwise the line $L''$ about $\bar a$, we obtain a valid line $L\in E(A(u,v,w,x))$.

In the case when $\bar a$ does not belong to the convex hull of $u,v,w$, there are among $u,v,w$ a rightmost point $z_r$ with abscissa $\bar a_1$  and a different lowest point $z_\ell$ with ordinate $\bar a_2$. The line through $z_r$ and $z_\ell$ has positive slope and all the points $a=(a_1,a_2)$ in the segment $z_rz_\ell$ belong to the convex hull of  $u,v,w$, so $x_1>a_1$ or $x_2<a_2$.
In particular, if $x_1>\bar a_1$, then $x_1>a_1$ for every point in the segment $z_rz_\ell$. So, we can take as $L'$ the vertical line through $z_r$. As before, slightly rotating $L'$ about $z^r$ clockwise we get a line $L$ as desired. Otherwise, if $x_2<\bar a_2$, then $x_2<a_2$ for every point in the segment $z_rz_\ell$. So, we can take as $L''$ the horizontal line through $z_\ell$. As before slightly rotating $L''$ about $z^\ell$ counterclockwise we get a line $L$ as desired. Finally, if $x_1\le \bar a_1$ and $x_2\ge \bar a_2$, let us consider the line $L$ through $z^r$ and $z^\ell$ and show that it belongs to $E(A(u,v,w,x))$. Indeed, by construction, $L$ has positive slope, the points $u,v,w$ are to the left of or on $L$ so they push right onto $L$ and $x$ is strictly on the right of $L$ because $x_1\le \bar a_1$, $x_2\ge \bar a_2$, and, for all the points $a=(a_1,a_2)$ of the segment $z_rz_\ell$, $x_1>a_1$ or $x_2<a_2$. Hence, the line $L$ belongs to $E(A(u,v,w,x))$.

It remains to prove the third statement (the fourth being similar). The proof is similar to that of the first statement.

Let us assume there is a point $a=(a_1,a_2)$  in the convex hull of $u,v,w$  such that $x_1\le  a_1$, $x_2\ge a_2$, and $x\ne a$, i.e. $x_1<a_1$ and $x_2>a_2$. As $a$ is in the convex hull of $u,v,w$, if there were a  line $L:y_2=my_1+q$ with slope $m>0$ separating $x$ from $u,v,w$, then $L$ would also separate $x$ from $a$. So, $a_2\ge ma_1+q$ while $x_2\le mx_1+q$. Now, the assumptions $x_1< a_1$ and $x_2> a_2$ imply $x_2> mx_1+q$,  a contradiction. 

To prove the opposite implication for statement 3, we notice that we may use an identical argument as in statement 1, except that strict inequalities become inclusive inequalities.  However, all of the results still hold, so that the opposite implication in statement 3 holds.

\end{proof}


 \begin{proof}[Proof of \Cref{prop:omega-C}]
 We will only prove case {\em 1}, as case {\em 2} is completely similar. 
 
 Let us first assume that $ E(A(u,v,w,x))$ contains a line of positive slope $L$ with $\Delta_L(u, v, w, x; \delta,\eta)=0$. By \cite[Lemma 4.1]{bapat2022computing}, such line $L$ must pass through a point $\omega=(\omega_1,\omega_2)$ determined by either \cref{eq:omega-1}  or  \cref{eq:omega-3}. So, also $ E(A(u,v,w,\omega))$ with $A_\omega=\{1,2\}$ is non-empty.
  In either \cref{eq:omega-1,eq:omega-3}, $\omega_1=x_1$, so that the assumption $A_x=\{1\}$, equivalent to $x\notin L$ and $x$ pushing upwards, implies that $\omega_2>x_2$.
 %
 Moreover, the third statement of \Cref{prop:erroneous-config-3vs1} applied to the points $u,v,w,\omega$, implies that there is no point $a=(a_1,a_2)$  in the convex hull of $u,v,w$ for which $\omega_1\le a_1$, $\omega_2\ge a_2$, and $\omega\ne a$. 
 
 
Vice versa,  we must show that among the lines $L$ of $E(A(u,v,w,x))$, which by assumption is non-empty, there is at least one line for which  $\Delta_L(u, v, w, x; \delta,\eta)=0$.
As $\omega$ is obtained by  \cref{eq:omega-1} or  \cref{eq:omega-3}, we have $\omega_1=x_1$.
Because we are assuming that there is no point $a=(a_1,a_2)$   in the convex hull of $u,v,w$ with $\omega_1\le a_1$, $\omega_2\ge a_2$, and $\omega\ne a$, the third statement of \Cref{prop:erroneous-config-3vs1} implies that $E(A(u,v,w,\omega))$ with $2\in A_u\cap A_v\cap A_w$ and $A_\omega=\{1,2\}$, is non-empty. Let $L'$ be a line in  $E(A(u,v,w,\omega))$. 
From the fact that $L'$ passes through $\omega$, $\omega_1=x_1$, and we are assuming that $\omega_2>x_2$, it follows that $A_x^{L'}=\{1\}$. So, $L'\in E(A(u,v,w,x))$.

Recall that  \cref{eq:omega-1} (resp. \cref{eq:omega-3}) for $\omega$ are obtained by imposing $\Delta_L(u, v, w, x; \delta,\eta)=0$ for $L\in E(A(u,v,w,x))$, implying that $\Delta_{L'}((u, v, w, x; \delta,\eta)=0$. 
 
 \end{proof}


 \begin{proof}[Proof of \Cref{prop:omega-D}]
We only prove the first equivalence, the second one being analogous. Let us consider the sets
$$A:=\{a\in \RR^2\mid a\in \mathrm{convhull}(u,v,w),\omega_1\leq a_1, \omega_2\geq a_2, \textrm{ and } a\neq\omega\}$$
and
$$B:=\{a\in \RR^2\mid a\in \mathrm{convhull}(u,v,w),\omega_1\leq a_1, \omega_2\geq a_2\}.$$
Clearly, $B$ contains $A$ and either $B=A\cup\{\omega\}$ or $B=A$.  Thus,  $A=\emptyset$ if and only if $B=\emptyset$ or $B=\{\omega\}$. In the first case, \Cref{alg:checkpts} can be used on the quadruple $u,v,w,\omega$ to determine if $B$ is empty: $B=\emptyset$ if and only if $\mathrm{CheckPts}(u,v,w,\omega, \mathrm{Below})=\mathrm{True}$. In the second case, $B=\{\omega\}$ if and only if $\omega$ is a point on the boundary of the convex hull of $u,v,w.$  In this case, \Cref{alg:checkpts} will return False for $u,v,w,\omega$.
\end{proof}


\begin{proof}[Proof of \Cref{prop:lubparametercheck}]
We know that $\lub(w,x)=x$ if and only if $x_1\geq w_1$ and $x_2\geq w_2$.  In this case, it must also be true that $\push_L(x)_1\geq \push_L(w)_1$ and $\push_L(x)_2\geq\push_L(w)_2$. By \cite[Equation (5)]{bapat2022computing}, this ensures that $p_L(x)\geq p_L(w)$. If $\lub(w,x)=w$, the above inequalities are swapped, and we obtain $p_L(x)\leq p_L(w)$.

If $\lub(w,x)$ is neither $x$ nor $w$, then the sign of $p_L(x)-p_L(w)$ depends on how $L$ separates $w,x,$ and $\lub(w,x)$. 
If $L$ separates $x$ and $\lub(w,x)$ from $w$, then $\push_L(x)=\push_L(\lub(w,x))$.  It is again the case that $\lub(w,x)_1\geq w_1$ and $\lub(w,x)_2\geq w_2$, so that $\push_L(\lub(w,x))_1\geq \push_L(w)_1$ and $\push_L(\lub(w,x))_2\geq\push_L(w)_2$.  Since we have $\push_L(x)=\push_L(\lub(w,x))$, we may again use \cite[Equation (5)]{bapat2022computing} to see that $p_L(x)\geq p_L(w)$.
If $L$ separates $w$ and $\lub(w,x)$ from $x$, then the above inequalities are swapped, and we obtain $p_L(x)\leq p_L(w)$.

\end{proof}


\begin{proof}[Proof of \Cref{prop:erroneous-config-2vs2pair}]
\underline{$(iii)$ implies $(ii)$:} 

Assume that there are a point $a=(a_1,a_2)$  in the convex hull of $u,v$ and a point $b=(b_1,b_2)$ in the convex hull of $x,w$ such that $(b_1\ge  a_1)\wedge (b_2\le a_2)$. If there is a line $L:y_2=my_1+q$ which separates $u,v$ from $w,x$ with $u,v$ pushing right and $x,w$ pushing up, then this line also strictly separates $a$ from $b$, with $a$ pushing right, and $b$ pushing up to $L$.  So, $a_2>ma_1+q$ and $b_2<mb_1+q$.  However, the assumption that $(b_1\le  a_1)\wedge (b_2\ge a_2)$ implies that 
    \[
    b_2<mb_1+q\leq ma_1+q<a_2\leq b_2,
    \]

\noindent which is a contradiction. Hence, the set of lines $E(A(u,v,w,x))$ is empty. 

\underline{$(ii)$ implies $(iii)$:} 

We show the contrapositive statement: If, for every point $a=(a_1,a_2)$ on $\overline{uv}$ and every point $b=(b_1,b_2)$ on $\overline{xw}$, we have that 
\begin{align}
    (b_1> a_1)\vee (b_2< a_2) \, ,
    \label{eq:assumption}
\end{align} then there exists a line $L$ in the set $E(A(u,v,w,x))$, i.e.~a line with positive slope that separates $u$ and $v$ from $x$ and $w$ with $ A^L_u= A^L_v=\{2\}$ and $A^L_x=A^L_w=\{1\}$. 

We start noting that a line $L: y_2 = my_1+q$ separates $u$ and $v$ from $x$ and $w$ with $ A^L_u= A^L_v=\{2\}$ and $A^L_x=A^L_w=\{1\}$ exactly when, for all $a=(a_1,a_2) \in \overline{uv}$ and all $b=(b_1,b_2) \in \overline{xw}$, the parameters $m$ and $q$ satisfy the following system of inequalities:

\begin{equation}\label{findline}
    \begin{cases}
      a_2 > ma_1 + q \\ 
        b_2 < mb_1 + q \,,
    \end{cases}
\end{equation}
or equivalently $a_2-ma_1 > q > b_2-mb_1 \, .$

For a given pair of points $a,b$, this system can be solved for $m$ and $q$ if and only if there exists an $m$ such that $a_2-ma_1 > b_2-mb_1$, or equivalently $a_2 - b_2 > m(a_1 - b_1)$. So there exists a line with positive slope that separates 
$a$ from $b$ with $A^L_a = \{2\}$ and $A^L_b = \{1\}$ if and only if there is a positive $m$ such that $a_2-b_2 > m(a_1-b_1)$. But this is the case exactly when we have $(b_1> a_1)\vee (b_2< a_2)$,  the assumption we are making for any $a=(a_1,a_2) \in \overline{uv}$ and $b=(b_1,b_2) \in \overline{xw}$. Therefore, we now know that, for a given pair of points $a=(a_1,a_2) \in \overline{uv}$ and $b=(b_1,b_2) \in \overline{xw}$, we can always find a line $L: y_2 = my_1+q$ with positive slope that separates $a$ from $b$ with $A^L_a = \{2\}$ and $A^L_b = \{1\}$.

Next, note that for a given pair of points $a=(a_1,a_2)$ and $b=(b_1,b_2)$ that satisfies the condition that $(b_1 >  a_1)\vee(b_2 < a_2)$, we have exactly one of the following three scenarios:
\begin{itemize}
    \item \underline{Scenario I}: $b_1 > a_1$ and $b_2 > a_2$
    \item \underline{Scenario II}: $b_1 > a_1$ and $b_2 < a_2$ 
    \item \underline{Scenario III}: $b_1 < a_1$ and $b_2 < a_2$
\end{itemize}
The scenario when  $b_1 < a_1$ and $b_2 > a_2$ is excluded by the assumption. 
To find line parameters $m$ and $q$ that solve the system of inequalities~(\ref{findline}) for a given pair of points $a=(a_1,a_2)$ and $b=(b_1,b_2)$, it may be helpful to visualise the lines $q=-b_1m +b_2$ and $q=-a_1m +a_2$ in the $(m,q)$-parameter space (see \Cref{m-q-space}). Note that in scenarios I and II, one can choose $m$ arbitrarily large, while in scenarios II and III, one can choose $m$ to be a positive number arbitrarily close to $0$. 
\begin{figure}[h]
    \includegraphics[scale=.4]{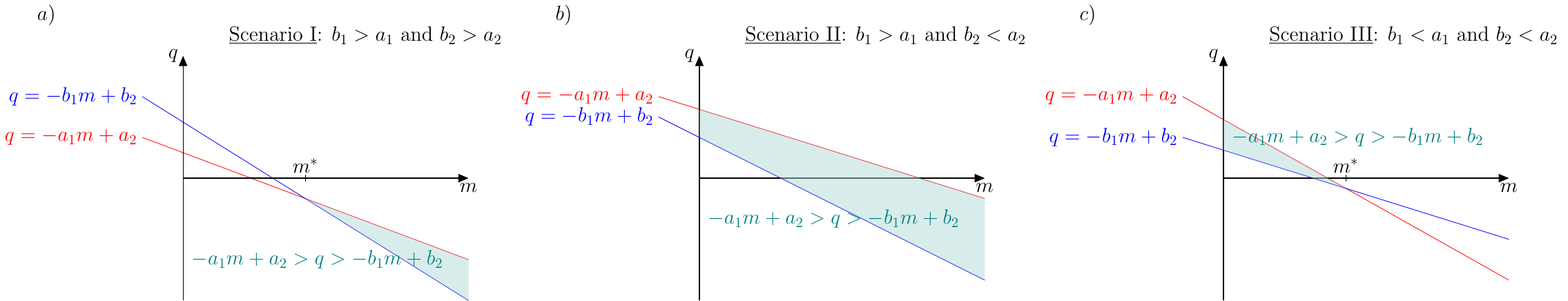} 
\caption{The lines $q=-a_1m +a_2$ (red) and $q=-b_1m +b_2$ (blue) and the regions in the $(m,q)$-parameter space corresponding to lines of positive slope that separate $a=(a_1,a_2)$ from $b=(b_1,b_2)$ with $A^L_a=\{2\}$ and $A^L_b=\{1\}$ (shaded in teal).} \label{m-q-space}
\end{figure}

To find a line of positive slope that separates $u$ and $v$ from $x$ and $w$ with $A^L_u= A^L_v=\{2\}$ and $A^L_x=A^L_w=\{1\}$, we need to find a positive $m$ and a $q$ that solve \ref{findline} for all $a=(a_1,a_2)\in \overline{uv}$ and $b=(b_1,b_2)\in\overline{xw}$. We will show below that there is always one endpoint of the segment $\overline{uv}$, one endpoint of the segment $\overline{xw}$, and a line with positive slope that separates these endpoints from each other in the desired way and whose slope guarantees that it in fact separates all of $\overline{uv}$ from all of $\overline{xw}$ in the desired way. The existence of such a line proves that $E(A(u,v,w,x))$ is non-empty. 

First, we define
\begin{equation}\label{slope_a}
m_a := \left\{\begin{array}{lr}
    \text{the slope of the line segment }\overline{uv} &\text{if } u \neq v\\
    0 &\text{if } u=v \ ,
    \end{array} \right.
\end{equation}
and 
\begin{equation}\label{slope_b}
m_b := \left\{\begin{array}{lr}
    \text{the slope of the line segment }\overline{xw} &\text{if } x \neq w\\
    0 &\text{if } x=w \ .
    \end{array} \right.
\end{equation}

We also define the following notation for the endpoints of the line segments $\overline{uv}$ and $\overline{uv}$:  
\begin{itemize}
\item Let $\{a',a''\}=\{u,v\}$ such that either $a''_1<a'_1$ or $(a'_1 = a''_1) \wedge (a''_2\leq a'_2)$ or $(a'_1 = a''_1) \wedge (a''_2 = a'_2)$, and
\item let $\{b',b''\}=\{x,w\}$ such that either $b'_1<b''_1$ or $(b'_1 = b''_1) \wedge (b'_2\leq b''_2)$ or $(b'_1 = b''_1) \wedge (b''_2 = b'_2) \ .$
\end{itemize}

With this notation, we can observe that the following four statements hold. 

\begin{enumerate}[label=(\subscript{S}{{\arabic*}})]\label{statements}
    \item Any line $L$ with positive slope $m \geq m_a$ and with $A^L_{a'}=\{2\}$ is such that $A^L_a=\{2\}$ for any $a\in \overline{uv}$. 
    \item Any line $L$ with positive slope $m \leq m_a$ and with $A^L_{a''}=\{2\}$ is such that $A^L_a=\{2\}$ for any $a\in \overline{uv}$. 
    \item Any line $L$ with positive slope $m \geq m_b$ and with $A^L_{b'}=\{1\}$ is such that $A^L_b=\{1\}$ for any $b\in \overline{xw}$.
    \item Any line $L$ with positive slope $m \leq m_b$ and with $A^L_{b''}=\{1\}$ is such that $A^L_b=\{1\}$ for any $b\in \overline{xw}$.

\end{enumerate}

Given $u,v,x$ and $w$, we can distinguish four cases based on the sign of the slopes of the line segments $\overline{uv}$ and $\overline{xw}$ (see Figure~\ref{cases_illustration}).\\

\begin{figure}[htb!]
    \begin{center}
    \begin{subfigure}{0.45\textwidth}
        \begin{center}
        		\begin{tikzpicture}
				\begin{tikzpicture}[]
\node[anchor=west] at (-1,2) {\underline{Case 1}: $m_a \leq 0$ and $m_b \leq 0$};
    \coordinate (1_a2) at (0,1);
    \coordinate (1_a1) at (.5,0);
    \coordinate (1_b1) at (1.5, 1);
    \coordinate (1_b2) at (2,0);
    \draw[red] (1_a1)--(1_a2);
    \draw[blue] (1_b1)--(1_b2);
    \node[red, anchor=north] at (1_a1) { $a'$};
     \node[red, anchor=south] at (1_a2) { $a''$};
    \node at (1_a1)[ circle,fill,red,inner sep=.8pt]{};
     \node[blue, anchor=south] at (1_b1) { $b'$};
     \node[blue, anchor=north] at (1_b2) { $b''$};
    \node at (1_b1)[ circle,fill,blue,inner sep=.8pt]{};

\end{tikzpicture}
			\end{tikzpicture}
        \end{center}
  \end{subfigure}
 \hfill
     \begin{subfigure}{0.45\textwidth}
        \begin{center}
        		\begin{tikzpicture}
				\begin{tikzpicture}[]
\node[anchor=west] at (-1,2) {\underline{Case 2}: $m_a \leq 0$ and $m_b > 0$};
    \coordinate (1_a2) at (0,1);
    \coordinate (1_a1) at (.5,0);
    \coordinate (1_b1) at (1.5, 0);
    \coordinate (1_b2) at (2,1);
    \draw[red] (1_a1)--(1_a2);
    \draw[blue] (1_b1)--(1_b2);
    \node[red, anchor=north] at (1_a1) { $a'$};
     \node[red, anchor=south] at (1_a2) { $a''$};
    \node at (1_a1)[ circle,fill,red,inner sep=.8pt]{};
     \node[blue, anchor=north] at (1_b1) { $b'$};
     \node[blue, anchor=south] at (1_b2) { $b''$};
    \node at (1_b1)[ circle,fill,blue,inner sep=.8pt]{};
    \node at (1_b2)[ circle,fill,blue,inner sep=.8pt]{};

\end{tikzpicture}
			\end{tikzpicture}
        \end{center}
  \end{subfigure}
\end{center}
    \begin{center}
\begin{subfigure}{0.45\textwidth}
        \begin{center}
        		\begin{tikzpicture}
				\begin{tikzpicture}[]
\node[anchor=west] at (-1,2) {\underline{Case 3}: $m_a > 0$ and $m_b \leq 0$};
    \coordinate (1_a2) at (0,0);
    \coordinate (1_a1) at (.5,1);
    \coordinate (1_b1) at (1.5, 1);
    \coordinate (1_b2) at (2,0);
    \draw[red] (1_a1)--(1_a2);
    \draw[blue] (1_b1)--(1_b2);
    \node[red, anchor=south] at (1_a1) { $a'$};
     \node[red, anchor=north] at (1_a2) { $a''$};
    \node at (1_a1)[ circle,fill,red,inner sep=.8pt]{};
    \node at (1_a2)[ circle,fill,red,inner sep=.8pt]{};
     \node[blue, anchor=south] at (1_b1) { $b'$};
     \node[blue, anchor=north] at (1_b2) { $b''$};
    \node at (1_b1)[ circle,fill,blue,inner sep=.8pt]{};

\end{tikzpicture}
			\end{tikzpicture}
        \end{center}
  \end{subfigure}
 \hfill
    \begin{subfigure}{0.45\textwidth}
        \begin{center}
        		\begin{tikzpicture}
				\begin{tikzpicture}[]
\node[anchor=west] at (-1,2) {\underline{Case 4}: $m_a > 0$ and $m_b > 0$};
    \coordinate (1_a2) at (0,0);
    \coordinate (1_a1) at (.5,1);
     \coordinate (1_b1) at (1.5, 0);
    \coordinate (1_b2) at (2,1);
    \draw[red] (1_a1)--(1_a2);
    \draw[blue] (1_b1)--(1_b2);
    \node[red, anchor=south] at (1_a1) { $a'$};
     \node[red, anchor=north] at (1_a2) { $a''$};
    \node at (1_a1)[ circle,fill,red,inner sep=.8pt]{};
    \node at (1_a2)[ circle,fill,red,inner sep=.8pt]{};
     \node[blue, anchor=north] at (1_b1) { $b'$};
     \node[blue, anchor=south] at (1_b2) { $b''$};
    \node at (1_b1)[ circle,fill,blue,inner sep=.8pt]{};
    \node at (1_b2)[ circle,fill,blue,inner sep=.8pt]{};

\end{tikzpicture}
			\end{tikzpicture}
        \end{center}
  \end{subfigure}
\end{center}
\caption{Illustration of slopes of the line segments $\overline{uv}$ and $\overline{xw}$ in the four different cases we consider. Note that these illustrations do not reflect all possible constellations. In particular, they do not reflect the cases where either $u=v$ or $x=w$.}
\label{cases_illustration}
\end{figure}
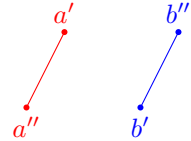

In the case of both $m_a$ and $m_b$ being positive, we distinguish further two subcases:
    \begin{itemize}
        \item Case 4a): $m_a \geq m_b$ 
        \item Case 4b): $m_a < m_b$ 
    \end{itemize}

In each of these cases, we next show that there is an endpoint of the segment $\overline{uv}$ and an endpoint of the segment $\overline{xw}$ and a line $L$ with positive slope separating these two endpoints in the desired way which in fact separates all of $\overline{uv}$ from all of $\overline{xw}$ in the desired way, that is with $A^L_u= A^L_v=\{2\}$ and $A^L_x=A^L_w=\{1\}$.

\underline{Case 1:} $m_a \leq 0$ and $m_b \leq 0$

 As observed above, by applying the assumption of \Cref{eq:assumption} to  ${a'}$ to ${b'}$ (i.e. $(b'_1 >  a'_1)\vee(b'_2 < a'_2)$), we can find a line $L$ of positive slope $m$ with $A^L_{a'}=\{2\}$ and $A^L_{b'}=\{1\}$. As any positive slope $m$ is greater than $m_a$ and $m_b$,  in this case, the line $L$ is such that $A^L_a=\{2\}$ for all $a \in \overline{uv}$ by statement~$(S_1)$ above, and such that $A^L_b=\{1\}$ for all $b \in \overline{xw}$ by statement~$(S_3)$ above. In other words, $L \in E(A(u,v,w,x))$.

\underline{Case 2:} $m_a \leq 0$ and $m_b > 0$. 

 If ${a'}$ and ${b'}$ give scenario I or II, then we can find a line $L$ with positive slope $m \geq m_b>0\geq m_a$ and with $A^L_{a'}=\{2\}$ and $A^L_{b'}=\{1\}$, as the range of possible slopes is not bounded above in these scenarios (cf. \Cref{m-q-space}(a-b). By statements $(S_1)$ and $(S_3)$ above, $L$ is such that $A^L_a=\{2\}$ for all $a \in \overline{uv}$ and such that $A^L_b=\{1\}$ for all $b \in \overline{xw}$. 

Similarly, if ${a'}$ and ${b''}$ give scenario II or III, we can find a line $L$ with positive slope $m$ such that $m_a<m \leq m_b$ and with $A^L_{a'}=\{2\}$ and $A^L_{b''}=\{1\}$, and this line $L$ is such that $A^L_a=\{2\}$ for all $a \in \overline{uv}$ and such that $A^L_b=\{1\}$ for all $b \in \overline{xw}$ by statements $(S_1)$ and $(S_4)$ above. 

Finally, if ${a'}$ and ${b'}$ give scenario III while ${a'}$ and ${b''}$ give scenario I, we first note that the following inequalities hold: 
\begin{equation}
\begin{split}
    &b_1' < a_1' < b_1'', \\ &b_2'< a_2' < b_2'' \ .
\end{split}
\label{eq:case3}
\end{equation} 
These inequalities together with the condition on the relative position of $a'$ to any point $b \in \overline{xw}$, i.e. $({b}_1 >  a'_1)\vee({b}_2 < a'_2)$, imply that the point $a'$ lies in the interior of the triangle with vertices $(b_1',b_2'),(b_1'',b_2'')$ and $(b_1',b_2'')$ and that the slope of the line segment $\overline{a'b'}$ is larger than $m_b$ (cf.~\Cref{rec_graph}a)). Now note that the $m$-coordinate of the intersection between the  lines $q=-b'_1m +b'_2$ and $q=-a'_1m +a'_2$ in the $(m,q)$-parameter space  is $m^* = \frac{a_2'-b_2'}{a_1'-b_1'}$,  i.e.~the slope of the line segment $\overline{a'b'}$. So by our observation above, we know that $m^*>m_b$, and, as ${a'}$ and ${b'}$ give scenario III, we can see from \Cref{m-q-space}c) that there exists a line $L$ of positive slope $m\geq m_b$ that separates $a'$ from $b'$ with $A^{L}_{a'}=\{2\}$, $A^{L}_{b'}=\{1\}$. Thus, by statements $(S_1)$ and $(S_3)$ we have that $L \in E(A(u,v,w,x))$. Note that one can argue similarly to show that there exists a line $L'$ of positive slope $m'\leq m_b$ that separates $a'$ from $b''$ with $A^{L'}_{a'}=\{2\}$, $A^{L'}_{b''}=\{1\}$, and that we have that $L' \in E(A(u,v,w,x))$ by statements $(S_1)$ and $(S_4)$. 

\begin{figure}[h]
     \begin{center}     
    \includegraphics[scale=1]{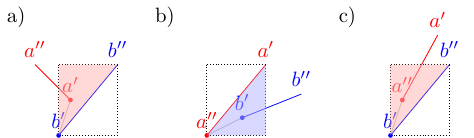} 
     \end{center}
\caption{Illustration of constellations of the points $a',a'',b'$ and $b''$ in subcases of case 2) and 4a). Panel a) shows the relative position of $a'$ with respect to $\overline{b'b''}$ when $a'$ and $b'$ give scenario III while $a'$ and $b''$ give scenario I in case 2. Panel b) shows the relative position of $b'$ with respect to $\overline{a''a'}$ when $a''$ and $b'$ give scenario I, and panel c) shows the relative position of $a''$ with respect to $\overline{b'b''}$ when $a''$ and $b'$ give scenario III. 
}
\label{rec_graph}
\end{figure}

\underline{Case 3:} $m_a > 0$ and $m_b \leq 0$.  

This case is analogous to Case~2. There is a line $L$ with positive $m \leq m_a$ such that $A^L_{a''}=\{2\}$, $A^L_{b'}=\{1\}$, and by statements $(S_2)$ and $(S_3)$ we have that $L \in E(A(u,v,w,x))$, and there is a line $L'$ with slope $m' \geq m_a$ such that $A^{L'}_{a'}=\{2\}$, $A^{L'}_{b'}=\{1\}$, and by statements $(S_1)$ and $(S_3)$ we have that $L' \in E(A(u,v,w,x))$. 

\underline{Case 4:} $m_a > 0$ and $m_b > 0$. 

If ${a''}$ and ${b''}$ give scenario II or III, then we can find a line $L$ with positive slope $m \leq m_a,m_b$ such that  $A^L_{a''}=\{2\}$,  $A^L_{b''}=\{1\}$, and by statements $(S_2)$ and $(S_4)$ we have that $L \in E(A(u,v,w,x))$ (as $m$ can be chosen arbitrarily close to $0$ in these scenarios).

 If ${a'}$ and ${b'}$ give scenario I or II, then we can find a line $L$ with slope $m \geq m_a,m_b$ such that $A^L_{a'}=\{2\}$, $A^L_{b'}=\{1\}$, and by statements $(S_1)$ and $(S_3)$ we have that $L \in E(A(u,v,w,x))$ (as $m$ can be chosen arbitrarily large in these scenarios). \ 
 
 If ${a''}$ and ${b''}$ give scenario I and ${a'}$ and ${b'}$ give scenario III, we first note that the following inequalities hold: 
\begin{equation}
\begin{split}
    a_1'' < b_1'', \quad  a_2'' < b_2'', \quad
a_1' > b_1', \quad  a_2' > b_2' \ .
\end{split}
\label{eq:case4last}
\end{equation}

We now consider the subcases 4a) and 4b) separately: 

\underline{Case 4a):} $m_a \geq m_b > 0$. We distinguish three further subcases based on the relative position of $a''$ to $b'$: 
\begin{itemize}
    \item $a''$ and $b'$ give scenario I, 
    \item $a''$ and $b'$ give scenario II, and
    \item $a''$ and $b'$ give scenario III.
\end{itemize}

In the case where $a''$ and $b'$ give scenario I, we have that 
\begin{equation}
\begin{split}
    a_1'' < b_1', \text{ and } a_2'' < b_2' \ .
\end{split}
\label{eq:case4apart1}
\end{equation}
Combining \Cref{eq:case4apart1} and \Cref{eq:case4last} gives
\begin{equation}
\begin{split}
    a_1'' < b_1' < a_1' \text{ and } a_2'' < b_2' < a_2'\ ,
\end{split}
\end{equation} 
and together with our assumption on the relative position of $b'$ to any point $a \in \overline{uv}$, namely that $(b'_1 >  a_1)\vee(b'_2 < a_2)$, this implies that $b'$ lies in the interior of the triangle with vertices $(a_1'',a_2''),(a_1',a_2'')$ and $(a_1',a_2')$, which in turn implies that the slope of the line segment $\overline{a''b'}$ is smaller than $m_a$ (cf.~\Cref{rec_graph}b)). But this guarantees the existence of a line $L$ separating $a''$ from $b'$ with $ A^L_{a''}=\{2\}$ and $A^L_{b'}=\{1\}$ whose slope $m$ is such that $m_b \leq m \leq m_a$: The $m$-coordinate of the intersection between the  lines $q=-b'_1m +b'_2$ and $q=-a''_1m +a''_2$ in the $(m,q)$-parameter space is $m^* = \frac{a_2''-b_2'}{a_1''-b_1'}$, i.e.~the slope of the line segment $\overline{a''b'}$. So by our observation above, we know that $m^*<m_a$, and, as $a''$ and $b'$ are in scenario I, we can see from \Cref{m-q-space}a) that there exists a line $L$ separating $a''$ from $b'$ with $ A^L_{a''}=\{2\}$ and $A^L_{b'}=\{1\}$ whose slope $m$ is such that $m_b \leq m \leq m_a$. 

If, on the other hand, $a''$ and $b'$ give scenario II, we can see in \Cref{m-q-space}b) that, for any positive $m$, there exists a line $L$ with slope $m$ that separates $a''$ from $b'$ with $ A^L_{a''}=\{2\}$ and $A^L_{b'}=\{1\}$. In particular, we can find a line $L$ separating $a''$ from $b'$ with $ A^L_{a''}=\{2\}$ and $A^L_{b'}=\{1\}$ whose slope $m$ is such that $m_b \leq m \leq m_a$. 

Finally, if $a''$ and $b'$ give scenario III, we have that 
\begin{equation}
\begin{split}
    a_1'' > b_1' \text{ and } a_2'' > b_2' \ .
\end{split}
\label{eq:case4apart3}
\end{equation}

Combining \Cref{eq:case4apart3} and \Cref{eq:case4last} gives
\begin{equation}
\begin{split}
    b_1' < a_1'' < b_1'' \text{ and } b_2' < a_2'' < b_2''\ ,
\end{split}
\end{equation} 
and together with our assumption on the relative position of $a''$ to any point $b \in \overline{xw}$, namely that $(b_1 >  a_1'')\vee(b_2 < a_2'')$, this implies that $a''$ lies in the interior of the triangle with vertices $(b_1',b_2'), (b_1'',b_2''),$ and $(b_1',b_2'')$, which in turn implies that the slope of the line segment $\overline{a''b'}$ is greater than $m_b$ 
(cf.~\Cref{rec_graph}c)). But this guarantees the existence of a line $L$ separating $a''$ from $b'$ with $ A^L_{a''}=\{2\}$ and $A^L_{b'}=\{1\}$ whose slope $m$ is such that $m_b \leq m \leq m_a$: Just as in scenario I, the $m$-coordinate of the intersection between the  lines $q=-b'_1m +b'_2$ and $q=-a''_1m +a''_2$ in the $(m,q)$-parameter space is $m^* = \frac{a_2''-b_2'}{a_1''-b_1'}$, i.e.~the slope of the line segment $\overline{a''b'}$. So by our observation above, we know that $m^*>m_b$, and, as $a''$ and $b'$ are in scenario III, we can see from \Cref{m-q-space}c) that there exists a line $L$ separating $a''$ from $b'$ with $ A^L_{a''}=\{2\}$ and $A^L_{b'}=\{1\}$ whose slope $m$ is such that $m_b \leq m \leq m_a$. 

So in any subcase, whether $a''$ and $b'$ give scenario I, II, or III, we know that there is a line $L$ separating $a''$ from $b'$ with $A^L_{a''}=\{2\}$ and $= A^L_{b'}=\{1\}$ whose slope $m$ is such that $m_b \leq m \leq m_a$. By statements $(S_2)$ and $(S_3)$ above, $L$ is such that $A^L_{a}=\{2\}$ and $= A^L_{b}=\{1\}$ for any $a \in \overline{uv}$ and $b \in \overline{xw}$. In other words, $L$ separates $u$ and $v$ from $x$ and $w$ in the desired way, and so $E(A(u,v,w,x))$ is non-empty.

\underline{Case 4b):} $m_b > m_a > 0$. 
This case is analogous to Case 4a). 
We can find a line $L$ with slope $m$ such that $m_a \leq m \leq m_b$ and with $A^L_{a'}=\{2\}$ and $A^L_{b''}=\{1\}$, and by statements $(S_1)$ and $(S_4)$ we have that $L$ is such that $A^L_{a}=\{2\}$ and $= A^L_{b}=\{1\}$ for any $a \in \overline{uv}$ and $b \in \overline{xw}$. In other words, $L \in E(A(u,v,w,x))$ and hence $E(A(u,v,w,x))$ is non-empty. 

In summary, we have found that in any case (Case 1, 2, 3, 4a), and 4b)) it is possible to find a line in the set $E(A(u,v,w,x))$, so this set is non-empty, and we have proven that not-$(iii)$ implies not-$(ii)$, and, equivalently, that $(ii)$ implies $(iii)$.

\underline{$(iv)$ implies $(iii)$:}

\noindent If it is the case that $x$ is in $Q$, then by definition of $Q$, $x_1\leq a_1$ and $x_2\geq a_2$ for some $a\in\overline{uv}$.  Thus, $x$ and $a$ satisfy the conditions given in (2).  If it is otherwise the case $w\in Q$ or $u$ or $v\in R$, similar pairs which satisfy the conditions of $(2)$ may be found with $w,u$ or $v$, respectively.

\underline{$(iii)$ implies $(i)$:}

\noindent Note that $\overline{uv}\in Q$ and $\overline{xw}\in R$.  If additionally we assume that there is some point $a=(a_1,a_2)$  on $\overline{uv}$ and a point $b=(b_1,b_2)$ on $\overline{xw}$ such that $(b_1\le  a_1)\wedge (b_2\ge a_2)$, then it is the case that $a\in R$ and $b\in Q$.  Thus, $a,b\in R\cap Q$, so clearly $R\cap Q\neq\emptyset.$

\underline{$(i)$ implies $(iv)$:}

\noindent To fix the ideas, we assume wlog $(u_2 > v_2) \vee \left((u_2=v_2) \wedge (u_1< v_1)\right)$, and $(w_1> x_1) \vee \left((w_1= x_1) \wedge (w_2< x_2)\right)$.

By definition of $Q$, if  $m_a\leq 0$ or $m_a=\infty$ (cf.~\Cref{slope_a}), then $Q=\{(p_1,p_2)\in\RR^2|p_1\leq v_1\textrm{ and }p_2\geq v_2\}$.  If  $\infty \neq m_a>0$, then $Q$ is the upper-region in $\RR^2$ bounded by 

\[\{(p_1,v_2)\in\RR^2|p_1\leq v_1\}\cup \overline{uv}\cup\{(u_1,p_2)\in\RR^2|p_2\geq u_2\}.\]

Similarly, if  $m_b\leq 0$ or $m_b=\infty$ (cf.~\Cref{slope_b}) , then $R=\{(p_1,p_2)\in\RR^2|p_1\geq w_1\textrm{ and }p_2\leq w_2\}$.  If  $\infty \neq m_a>0$, then $R$ is the lower-region in $\RR^2$ bounded by 

\[\{(x_1,p_2)\in\RR^2|p_2\leq x_2\}\cup \overline{xw}\cup\{(p_1,w_2)\in\RR^2|p_1\geq w_1\}.\]

\begin{figure}[htb]
    \begin{center}
    \begin{subfigure}{0.45\textwidth}
                \begin{center}
			\begin{tikzpicture}
				\input{./figures/2v2pairedUVposslope.tex}
			\end{tikzpicture}
                \end{center}
    \caption{ The region $Q$, when $\overline{uv}$ has positive slope, is shaded in red. }
    \label{fig:R-Q-regions-a}
    \end{subfigure}
 \hfill
        \begin{subfigure}{0.45\textwidth}
        \begin{center}
        		\begin{tikzpicture}
				\input{./figures/2v2pairedUVnegslope.tex}
			\end{tikzpicture}
        \end{center}
        \caption{The region $Q$, when $\overline{uv}$ has negative slope, is shaded in red.}
	\label{fig:R-Q-regions-b}
  \end{subfigure}
\end{center}

    \begin{center}
    \begin{subfigure}{0.45\textwidth}
         \begin{center}
			\begin{tikzpicture}
				\input{./figures/2v2pairedXWposslope.tex}
			\end{tikzpicture}
         \end{center}
    \caption{ The region $R$, when $\overline{xw}$ has positive slope, is shaded in blue. }
    \label{fig:R-Q-regions-c}
    \end{subfigure}
 \hfill
    \begin{subfigure}{0.45\textwidth}
        \begin{center}
        		\begin{tikzpicture}
				\input{./figures/2v2pairedXWnegslope.tex}
			\end{tikzpicture}
        \end{center}
    \caption{The region $R$, when $\overline{xw}$ has negative slope, is shaded in blue.}
    \label{fig:R-Q-regions-d}
  \end{subfigure}
\end{center}
\caption{The regions $Q$ and $R$ defined in \Cref{prop:erroneous-config-2vs2pair} }
\end{figure}

With this in mind, assume that $R\cap Q\neq\emptyset$. If it is the case that  $m_b \leq 0$ or $m_b=\infty$, then from the assumption $x_1\le w_1$ it follows that $x_2\ge w_2$, and thus we have $x\in Q$. The case when $m_a \leq 0$ or $m_a=\infty$ is analogous. 

If both $\infty \neq m_a>0$ and $\infty \neq m_b>0$, it must be the case that at least one of the boundary components of $R$ intersects $Q$ and vice versa.  

\begin{itemize}
    \item If $\{(x_1,p_2)\in\RR^2|p_2\leq x_2\}$ intersects $Q$, then $x\in Q$.
    \item If $\{(p_1,w_2)\in\RR^2|p_1\geq w_2\}$ intersects $Q$, then $w\in Q$.
    \item  If neither of the unbounded components of the boundary of $R$ intersects $Q$, it must be the case that at least one of the unbounded boundary components of $Q$ intersects $R$.  Using the same justification as in the previous two points, this means that either $u$ or $v$ is in $R$.
\end{itemize}

\end{proof}


\begin{proof}[Proof of \Cref{prop:omega-2pv2p}]
If the equivalence class $E(A(u,v,w,x))$ contains some lines $L$ for which  $\Delta_L(u, v, w, x; \delta,\eta)=0$, then by \cite[Lemma 4.1]{bapat2022computing} it passes through     $\omega$ as given in \cref{eq:omega-ter}. This implies that  $L$ has slope $m=\frac{\delta(w_1-x_1)}{\eta(u_2-v_2)}$, which is positive by the assumption  $p_L(u)>p_L(v)$ and $p_L(w)>p_L(x)$. As for the $y$-intercept $q$ of $L$,  because $L$ keeps $u,v$ on the left and $x,w$ on the right, there must exist $q\in\RR$ such that 
$\max\{x_2-mx_1,w_2-mw_1\}<q<\min\{u_2-m u_1,v_2-mv_1\}$, proving \Cref{eq:omega-2pv2p}. Vice versa,  we must show that among the lines $L$ of $E(A(u,v,w,x))$, which by assumption is non-empty, there is at least one line for which  $\Delta_L(u, v, w, x; \delta,\eta)=0$. 
If  $\max\{x_2-mx_1,w_2-mw_1\}<\min\{u_2-m u_1,v_2-mv_1\}$ then  taking $q$ between these two values, the line $L:y=mx+q$ is such that it passes through $\omega$ and correctly separates $u,v$ from $x,w$. As $\omega=\left[0:\delta(w_1-x_1):\eta(u_2-v_2)\right]$ has been obtained from \Cref{eq:omega-3} imposing $\Delta_L(u, v, w, x; \delta,\eta)=0$, we can conclude.

\end{proof}


\begin{proof}[Proof of \Cref{prop:omega-2uv2u}]
The first two cases are self-evident.  Suppose then that $x,v\notin Q_2$, $u,w\notin Q_4$, and at least one of $x,v,u,w$ is in $Q_1$ or $Q_3$. Notice that if $x\in Q_4$, then $x$ will push up to any line through $\omega$ with positive slope - the same is true for $v\in Q_4$.  Similarly, if $u\in Q_2$, then $u$ will push right to any line through $\omega$ with positive slope, and the same is true for $w\in Q_2$. So, we only need to consider the points $c$ in $\{x,v,u,w\}\cap (Q_1\cup Q_3)$, which by assumption is non-empty. If $c\in Q_1\cap P$, then a line $L$ through $\omega$  such that $A_c=\{1\}$ exists if and only if its slope $m$ is such that $m > m_c$, where $m_c$ is the slope through $\omega$ and $c$. If $c \in Q_3\cap P$, then the condition is $m < m_c$. Analogous conditions apply for $A_c=\{2\}$, $c\in Q_1\cap S$ and $c\in Q_3\cap S$.

\end{proof}

\section{Pseudo-code}\label{app:pseudo-code}
\addcontentsline{toc}{section}{Appendix}

\begin{algorithm}
\footnotesize{
  \caption{ Function to assess if, for four points $x, u,v,w$, there is a line $L$ with positive slope  separating $x$ from $u,v,w$.}\label{alg:checkpts}

\vspace{10pt}
\textbf{Input: }{$u,v,w,x$, Position} \\
\textbf{Output: }{ True if the configuration is good and False otherwise.}
\begin{algorithmic}[1]
\vspace{10pt}
\Function{CheckPts}{$u,v,w,x$, Position}\\

\If{Position == Below } \Comment{case $x$ strictly below the line and $u,v,w$ above the line } 
    \State{$z_1 \gets \max(u_1,v_1,w_1)$} 
    \State{$ z_2 \gets \min(u_2,v_2,w_2)$}
    \State{$z \gets (z_1, z_2)$}
    \State{\mbox{$Q \gets$  the closed upper-left quadrant with origin at $z$}} 
\EndIf \\

\If{Position == Above }\Comment{case $x$ strictly above the line and $u,v,w$ below the line }
   \State{$z_1 \gets \min(u_1,v_1,w_1)$} 
    \State{$ z_2 \gets \max(u_2,v_2,w_2)$}
    \State{$z \gets (z_1, z_2)$}
    \State{\mbox{$Q \gets$ the closed lower-right quadrant with origin at $z$} }
\EndIf\\

\If{$x \in Q$}
    \If{$x \in \textrm{convhull}(u,v,w)$}
    
        \State{ \Return{\Call{False}{}}}
        
    \ElsIf{$x \in \textrm{convhull}(u,v,w,z)$}
    
        \State{ \Return{\Call{True}{}}}
    \Else

      \State{ \Return{\Call{False}{}}}
    \EndIf
    
\Else

    \State{ \Return{\Call{True}{}}}

\EndIf
\EndFunction

\end{algorithmic}
}
\end{algorithm}

 \begin{algorithm}
\footnotesize{
  \caption{Functions to assess if $\omega$ satisfies the conditions of \Cref{prop:omega-C} and is not superfluous. }\label{alg:omega-C}
\textbf{Input: }{$\omega,u,v,w,x$, Position.}
\textbf{Output: }{ True, if $\omega$ is correct and not superfluous, otherwise False.}
\begin{algorithmic}[1]
\Function{CheckOmega}{$\omega,u,v,w,x$, Position}\\
    \If{Position == Below } \Comment{case $x$ strictly below the line and $u,v,w$ above the line }
        \If{$\omega_2\le x_2$ }
           \State{\Return{\Call{False}{}}}
        \EndIf\\
    \ElsIf{Position == Above } \Comment{case $x$ strictly above the line and $u,v,w$ below the line }
        \If{$\omega_1\le x_1$ }
           \State{\Return{\Call{False}{}}}
        \EndIf
    \EndIf\\

    \If{CheckPts({$u,v,w,\omega$, Position}) == False} \Comment{Both cases $x$ strictly above or below the line}
        \State{\Return{\Call{False}{}}}
    \Else
	\State{\Return{\Call{True}{}}}
    \EndIf
\EndFunction
\end{algorithmic}
}
\end{algorithm}


\begin{algorithm}
\scriptsize{
  \caption{An algorithm to compute switch points $\omega$ from a set of four points $c_1,c_2,c_3,c_4$ so that three points are one side of a line through $\omega$ and the remaining point is strictly on the other side.   }\label{alg:3vs1}

\textbf{Input: }{$c_1, c_2, c_3, c_4 \in C_M\sqcup C_N$.}
\textbf{Output: }{The partial list of switch points $\omega$ generated by Case 2
in \cite[Lemma 4.1]{bapat2022computing}}}
\begin{algorithmic}[1]

\Ensure{ there are 3 distinct points among  $\{c_1,c_2,c_3,c_4\} $ }
\Function{OmegaPoints3v1}{$c_1, c_2, c_3, c_4$} 
%
%
\State{ $\Omega\gets\emptyset$}
\State{CritPts $\gets \{c_1,c_2,c_3,c_4\} $}
\For{ $c$ in  CritPts } 

\If {\Call{check\_c\_distinct}{c, CritPts$\setminus$\{c\}}} 
\Comment{$x$ must be different from  $u,v,w$}
\State  {$x \gets c$}
\Else
\State { \bf{continue}}
\EndIf

\For{$d$ in CritPts.remove(x)}
\Comment{Note: $w$ therefore is distinct from $x$}
    \State{ $w \gets d$}
    
    \If{both elements of CritPts$\setminus$\{x,w\} are the same}{\bf{ continue}}
 	  \Comment{The crit vals assigned \\ \hfill{} to $u$ and $v$ cannot be the same.}
    \EndIf
        \State{ $u \gets$ an element of CritPts$\setminus$\{x,w\}}
        \State{ $v \gets $ CritPts$\setminus$\{x,w,u\}}

        \If{$x$.parent() == $w$.parent()} \Comment{determining $\eta$}
			  	
            \State{$\eta\gets 2$}
	\Else 
	  	\State{$\eta\gets 1$}
	\EndIf

        \If{$u$.parent() == $v$.parent()} \Comment{determining $\delta$}
	  	\State{$\delta\gets 2$}
	\Else 
	  	\State{$\delta\gets1$}
	\EndIf
   
        \Comment{Case when  $x$ pushes up, $u,v,w$ push right.}
        \If{\Call{CheckPts}{CritPts, Below}} \Comment{Checking if the configuration is possible with $x$ below the line \\ \hfill{} and $u,v,w$ above the line} 

            \If{$u_2 == v_2$}{\bf{ continue}}\Comment{This matching has null cost. }
            \EndIf
            \If{$u_2 < v_2$}\Comment{ensuring $p_L(u)>p_L(v)$}
                \State{ Exchange  $u,v$.}  
		\EndIf 
            \If{$\lub(x,w)=w$}\Comment{Case where $p_L(w)\ge  p_L(x)$}
   			\State{$\omega \gets (x_1,\frac{\eta}{\delta}(v_2- u_2) + w_2)$} \Comment{apply \Cref{eq:omega-1}}
 	 	\ElsIf{$\lub(w,x)=x$}\Comment{Case  $p_L(x)\ge p_L(w)$} 
 			\State{ $\omega \gets (x_1,\frac{\eta}{\delta}(u_2- 
v_2) + w_2)$ } \Comment{apply \Cref{eq:omega-3}}
            \Else\Comment{Case where $\lub(w,x)\neq x$ or $w$} 
 		    \State{ $\omega \gets (x_1,\frac{\eta}{\delta}(u_2- v_2) + w_2)$ }\Comment{apply \Cref{eq:omega-3}}
                \State{$\omega' \gets (x_1,\frac{\eta}{\delta}(v_2- u_2) + w_2)$} \Comment{apply \Cref{eq:omega-1}}
                 
                \If{\Call{CheckOmega}{$\omega',u,v,w,x$,Below}} \Comment{Check if  $\omega'$ satisfies condition 2 \\ \hfill{} of \Cref{prop:omega-C}}
                    \State{\Call{Append}{$\Omega, \omega'$}}
                \EndIf
            \EndIf   					 
         
            \If{\Call{CheckOmega}{$\omega,u,v,w,x$,Below}} \Comment{Checking if  $\omega$ satisfies condition 2  \\ \hfill{} of \Cref{prop:omega-C}}
                \State{\Call{Append}{$\Omega, \omega$}}
            \EndIf
        \EndIf
        
        \Comment{Case when  $x$ pushes right, $u,v,w$ push up}
        \If{\Call{CheckPts}{CritPts,Above}}\Comment{Checking if the configuration is possible with $x$ above the line \\ \hfill{} and $u,v,w$ below the line}

        \If{$u_1 == v_1$}{\bf{ continue}}\Comment{This matching has null cost. }
        \EndIf
		\If{$u_1 < v_1$} 
                \State{ Exchange $u,v$}	
		\EndIf 
            \Comment{ensuring $p_L(u)>p_L(v)$}
            \If{$\lub(w,x)=w$}\Comment{Case where $p_L(w)\ge p_L(x)$}
   	  				\State{ $\omega \gets (\frac{\eta}{\delta}(v_1- u_1) + w_1, x_2)$ }   \Comment{apply \Cref{eq:omega-2}}
 			 	\ElsIf{$\lub(w,x)=x$}\Comment{Case where $p_L(x)\ge p_L(w)$}
 					 \State{ $\omega \gets (\frac{\eta}{\delta}(u_1- v_1) + w_1, x_2)$ }\Comment{ apply \Cref{eq:omega-4}}
       \Else\Comment{Case where $\lub(w,x)\neq x$ or $w$} 
 				\State{ $\omega \gets (\frac{\eta}{\delta}(v_1- u_1) + w_1, x_2)$ }   \Comment{apply \Cref{eq:omega-2}}
   					 \State{ $\omega' \gets (\frac{\eta}{\delta}(u_1- v_1) + w_1, x_2)$ }\Comment{apply \Cref{eq:omega-4}}
                    
                    \If{\Call{CheckOmega}{$\omega',u,v,w,x$,Above}} \Comment{Checking if  $\omega'$ satisfies  condition 1 of \Cref{prop:omega-C}}              
                        \State{\Call{Append}{$\Omega, \omega'$}}
                    \EndIf
				\EndIf
     					     
                        \If{\Call{CheckOmega}{$\omega,u,v,w,x$,Above}} \Comment{Checking if  $\omega$ satisfies  condition 1 of \Cref{prop:omega-C}}          
                            \State{\Call{Append}{$\Omega, \omega$}}
                        \EndIf

            \EndIf
	\EndFor
	\EndFor 

\State{\Return{$\Omega$}}
\EndFunction
\vspace{10pt}
\end{algorithmic}

\end{algorithm}

\begin{algorithm}
\footnotesize{
  \caption{ Function to assess if, for distinct four points $x, u,v,w$, there is a line $L$ with positive slope  separating $x,w$ from $u,v$, where $x,w$ push up, and $u,v$ push right.}\label{alg:checkpts-2v2paired}

\vspace{10pt}
\textbf{Input: }{$u,v,w,x$} \\
\textbf{Output: }{ True if the configuration is good and False otherwise.}
\begin{algorithmic}[1]
\vspace{10pt}
\Function{CheckPts2}{$u,v,w,x$}\\

\State{Set $Q$ and $R$ as in \Cref{prop:erroneous-config-2vs2pair} }

\If{$x \in Q$ or $w\in Q$}
      \State{ \Return{\Call{False}{}}}
\ElsIf{$u \in R$ or $v\in R$}
      \State{ \Return{\Call{False}{}}}
\Else
    \State{ \Return{\Call{True}{}}}
\EndIf
\EndFunction

\end{algorithmic}
}
\end{algorithm}

\begin{algorithm}
\footnotesize{
  \caption{{\bf (2paired-vs-2paired)} An algorithm to compute switch points $\omega$ from a set of four points $c_1,c_2,c_3,c_4$ so that two paired points are one side of a with slope given by $\omega$ and the remaining two paired points are strictly on the other side. }\label{alg:2paired-vs-2paired}

\textbf{Input: }{$c_1, c_2, c_3, c_4 \in C_M\sqcup C_N$.}
\textbf{Output: }{The partial list of $\omega$ slopes generated by case 3
in the proof of \cite[Lemma 4.1]{bapat2022computing}}

\begin{algorithmic}[1]

\Ensure{ there are 4 distinct points among  $\{c_1,c_2,c_3,c_4\} $ }

\Function{OmegaSlopes2v2paired}{$c_1, c_2, c_3, c_4$} 
%
%
\State{ $\Omega\gets\emptyset$}
\State{CritPts $\gets \{c_1,c_2,c_3,c_4\} $}
\For{ $\{c,d\}$ in  CritPts } \Comment{Choose 2 of the 4 points}
    \State{$x \gets c$}
    \State{$w \gets d$}

     \If{$w_1 < x_1$}\Comment{ensuring $p_L(w)>p_L(x)$}
        \State{ Exchange  $w,x$} 
     \ElsIf{$w_1==x_1$}{ \bf{continue}} \Comment{This matching has null cost.}
      
	\EndIf 

    \State{ $u \gets $ an element of CritPts$\setminus\{x,w\}$}
    \State{ $v \gets $ CritPts$\setminus\{x,w,u\}$}
     
    \If{$u_2 < v_2$}\Comment{ensuring $p_L(u)>p_L(v)$}
        \State{ Exchange  $u,v$} 
     \ElsIf{$u_2==v_2$}{ \bf{continue}} \Comment{This matching has null cost.}
	\EndIf 
 
    \If{$x$.parent() == $w$.parent()} \Comment{determining $\eta$}
			  	
        \State{$\eta\gets 2$}
	\Else 
	  	\State{$\eta\gets 1$}
	\EndIf

    \If{$u$.parent() == $v$.parent()} \Comment{determining $\delta$}
	  	\State{$\delta\gets 2$}
	\Else 
	  	\State{$\delta\gets1$}
	\EndIf
   
        \If{\Call{CheckPts2}{$x,w,u,v$}} \Comment{Checking if the configuration is possible for labeled points} \\

        \State{$\omega \gets [0:\delta(w_1-x_1):\eta(u_2-v_2)]$} \Comment{Apply \Cref{eq:omega-ter}}
        \State{$m \gets \frac{\eta(u_2-v_2)}{\delta(w_1-x_1)}$}
        \If{$\max\{x_2-mx_1,w_2-mw_1\}<\min\{u_2-m u_1,v_2-mv_1\}$} 
        \Comment{Checking the condition of \Cref{prop:omega-2pv2p}}
            \State{\Call{Append}{$\Omega, \omega$}}
        \EndIf
    \EndIf
\EndFor 

\State{\Return{$\Omega$}}
\EndFunction
\end{algorithmic}
}
\end{algorithm}

 \begin{algorithm}
\footnotesize{
  \caption{Functions to assess if $\omega$ satisfies the conditions of \Cref{prop:omega-2uv2u} and is not superfluous.  }\label{alg:checkomega3}
\textbf{Input: }{$\omega,u,v,w,x$.}
\textbf{Output: }{ True, if $\omega$ is correct and not superfluous, otherwise False.}
\begin{algorithmic}[1]
\Function{CheckOmega2}{$\omega,u,v,w,x$}\\

    \For{$p\in\{u,v,w,x\}$}\Comment{Assign $\omega$ quadrants and associated slopes and signs for relevant points.}
        \If{$p_1>\omega_1$ and $p_2>\omega_2$}
            \State{$Q_p\gets 1$}
            \State{$m_p \gets \frac{p_2-\omega_2}{p_1-\omega_1}$}
            \If{$p\in\{v,x\}$}
                \State{$\textrm{sign}_p\gets 1$}
            \Else
                \State{$\textrm{sign}_p\gets -1$}
            \EndIf
        \ElsIf{$p_1\leq\omega_1$ and $p_2\geq\omega_2$}
            \State{$Q_p\gets 2$}
        \ElsIf{$p_1<\omega_1$ and $p_2<\omega_2$}
            \State{$Q_p\gets 3$}
            \State{$m_p \gets \frac{p_2-\omega_2}{p_1-\omega_1}$}
              \If{$p\in\{v,x\}$}
                \State{$\textrm{sign}_p\gets -1$}
            \Else
                \State{$\textrm{sign}_p\gets 1$}
            \EndIf
        \Else    
            \State{$Q_p\gets 4$}
        \EndIf    
    \EndFor
    \If{$\omega\in\{x,v,u,w\}$}
        \State{\Return{\Call{False}{}}}
    \ElsIf{$Q_x=2$ or $Q_v=2$ or $Q_u=4$ or $Q_w=4$}\Comment{Case 1 of \Cref{prop:omega-2uv2u}}
        \State{\Return{\Call{False}{}}}
    \ElsIf{$Q_x=4$ and $Q_v=4$ and $Q_u=2$ and $Q_w=2$ }\Comment{Case 2 of \Cref{prop:omega-2uv2u}}
        \State{\Return{\Call{True}{}}}
    \Else\Comment{Case 3 of \Cref{prop:omega-2uv2u}}
        \State{$P\gets\emptyset$, $S\gets\emptyset$}
        \For{$p\in\{v,x\}$}
            \If{$Q_p=4$}{\bf{ continue}}\Comment{Determine if $x$ or $v$ are in $Q_1\cup Q_3$}
            \Else
                \State{\Call{Append}{$P, p$}}
            \EndIf
        \EndFor
        \For{$p\in\{u,w\}$}
            \If{$Q_p=2$}{\bf{ continue}}\Comment{Determine if $u$ or $w$ are in $Q_1\cup Q_3$}
            \Else
                \State{\Call{Append}{$S, p$}}
                \EndIf
            \EndFor
            
        \For{$p\in P\cup S$} \Comment{Determine slope intervals for all points in $P\cup S$}
            \If{$\textrm{sign}_p=1$}
                \State{$\textrm{int}_p\gets (m_p,\infty)$}
            \Else
                \State{$\textrm{int}_p\gets (0,m_p)$}
            \EndIf
        \EndFor
        \If{$\displaystyle\cap_{p\in P}\textrm{int}_p\neq\emptyset$}\Comment{Determine existence of slope satisfying conditions of \Cref{prop:omega-2uv2u}}
            \State{\Return{\Call{True}{}}}
        \Else
            \State{\Return{\Call{False}{}}}
        \EndIf
    \EndIf
\EndFunction
\end{algorithmic}
}
\end{algorithm}
\begin{algorithm}
\footnotesize{
  \caption{{\bf (2unpaired-vs-2unpaired)} An algorithm to compute switch points $\omega$ from a set of four points $c_1,c_2,c_3,c_4$ so that two unpaired points are one side of a with slope given by $\omega$ and the remaining two unpaired points are strictly on the other side. }\label{alg:2unpaired-vs-2unpaired}

\textbf{Input: }{$c_1, c_2, c_3, c_4 \in C_M\sqcup C_N$.}
\textbf{Output: }{The partial list of $\omega$ points generated by case 4
in the proof of \cite[Lemma 4.1]{bapat2022computing}}

\begin{algorithmic}[1]
\Ensure{ there are 3 distinct points among  $\{c_1,c_2,c_3,c_4\} $ }
\Function{OmegaPts2v2unpaired}{$c_1, c_2, c_3, c_4$} 
    \State{ $\Omega\gets\emptyset$}
    \State{CritPts $\gets \{c_1,c_2,c_3,c_4\} $}
    \For{ $\{c,d\}$ in  CritPts } \Comment{Choose 2 of the 4 points} 
        \State{$x \gets c$}
        \State{$v\gets d$}
        \For{$f$ in  CritPts.remove($x,v$) } \Comment{take another point}
		  \If{$f==x$} { \bf{continue}} 
		  \Else  
    	   \State{$w \gets f$}
   	    \EndIf
   	    
            \For{$g$ in CritPts.remove($x,v,w$)} \Comment{take another point}
	  
                \If{$g==v$} {\bf{continue}} 
		      \Else  
    	       \State{$u \gets g$}
   	        \EndIf
      	 
                \If{$x$.parent() == $w$.parent()} \Comment{determining $\eta$}	  
                    \State{$\eta\gets 2$}
	        \Else 
                    \State{$\eta\gets 1$}
	        \EndIf

                \If{$u$.parent() == $v$.parent()} \Comment{determining $\delta$}
	  	    \State{$\delta\gets 2$}
	        \Else 
	  	    \State{$\delta\gets 1$}
	        \EndIf
   
                \If{\Call{CheckPts2}{$x,v,w,u$}} \Comment{Check if the configuration is possible} 
                   \If{$\left\{\begin{array}{ll}\left(\lub(u,v)=u \mbox{\ and\ } \lub(x,w)=w\right)& \mbox{\ or\ } \\
                   \left(\lub(u,v)=v \mbox{\ and\ } \lub(x,w)=x\right) & \mbox{\ or\ }\\ \lub(u,v)\notin\{u,v\} & \mbox{\ or\ } \\
                   \lub(x,w)\notin\{x,w\} &  \end{array}\right.$}
                    \If{$\delta=\eta$ and $v_1\ne x_1$ and $u_2\ne w_2$}
                    
       	                 \State{$\omega \gets [0:(v_1-x_1):(u_2-w_2)]$}        \Comment{Apply \Cref{eq:omega-quater}}
                            \State{$m \gets \frac{u_2-w_2}{v_1-x_1}$} \Comment{Check the condition of \Cref{prop:omega-2pv2p}}
                            \If{$m>0$ and $\max\{x_2-mx_1,v_2-mv_1\}<\min\{u_2-m u_1,w_2-mw_1\}$} 
                                \State{\Call{Append}{$\Omega, \omega$}}
                            \EndIf
		              \ElsIf{$\delta\ne \eta$}	               
                            \State{$\omega \gets \left(\frac{\eta v_1-\delta x_1}{\eta-\delta}, \frac{\eta u_2-\delta w_2}{\eta-\delta}\right)$} \Comment{Apply \Cref{eq:omega-5}}
                            \If{\Call{CheckOmega2}{$\omega,u,v,w,x$}}  \Comment{Check that some line through $\omega$ \\ \hfill{} gives $\Delta_L=0$}
                                \State{\Call{Append}{$\Omega, \omega$}}
                            \EndIf  
	                \EndIf
                \EndIf
               \If{$\left\{\begin{array}{ll}\left(\lub(u,v)=u \mbox{\ and\ } \lub(x,w)=x\right) & \mbox{\ or\ } \\
               \left(\lub(u,v)=v \mbox{\ and\ } \lub(x,w)==w\right) &\mbox{\ or\ }\\ \lub(u,v)\notin\{u,v\} &\mbox{\ or\ } \\
               \lub(x,w)\notin\{x,w\}\end{array}\right.$}
	               \State{$\omega \gets \left(\frac{\eta v_1+\delta x_1}{\eta+\delta}, \frac{\eta u_2+\delta w_2}{\eta+\delta}\right)$} \Comment{Apply \Cref{eq:omega-6}}

                       \If{\Call{CheckOmega2}{$\omega,u,v,w,x$}} \Comment{Check that some line through $\omega$ \\ \hfill{} gives $\Delta_L=0$}
                            \State{\Call{Append}{$\Omega, \omega$}}
                        \EndIf  
                    \EndIf
               
                \EndIf
            \EndFor
        \EndFor
    \EndFor 
    \State{\Return{$\Omega$}}
\EndFunction
\end{algorithmic}
}
\end{algorithm}

\end{document}